\documentclass[lettersize,journal]{IEEEtran}
\usepackage[margin=1.05in]{geometry}
\usepackage{textcomp}
\usepackage{enumitem}
\usepackage{stfloats}
\usepackage{url}
\usepackage[mathscr]{euscript}
\usepackage{cite}
\usepackage{import}
\usepackage{amsmath,amsfonts}
\usepackage{algorithmic}
\usepackage{array}
\usepackage[caption=false,font=normalsize,labelfont=sf,textfont=sf]{subfig}
\usepackage{textcomp}
\usepackage{stfloats}
\usepackage{url}
\usepackage{verbatim}
\usepackage{graphicx}
\usepackage{bm}
\usepackage{cite}
\usepackage{caption}
\usepackage{bbm}
\usepackage{wrapfig}
\usepackage[ruled,vlined]{algorithm2e}
\usepackage[dvipsnames]{xcolor}
\usepackage{setspace}
\usepackage{tikz}
\usetikzlibrary{shapes,arrows,positioning,calc}
\usetikzlibrary{arrows.meta,positioning, calc} 
\usetikzlibrary{matrix}
\usepackage{multirow}
\usepackage{pgfplots}
\pgfplotsset{compat=1.18}
\newcommand{\bld}[1]{\boldsymbol{#1}}

\newcommand{\sq}[1]{\left[#1\right]}
\newcommand{\cq}[1]{\left(#1\right)}
\newcommand{\bq}[1]{\left\{#1\right\}}

\newtheorem{theorem}{Theorem}

\def\maxs{MaxSINR~}
\def\maxsnospace{MaxSINR}
\def\disc{DISC-MaxSINR~}
\def\discnospace{DISC-MaxSINR}
\def\discpl{DISC-MaxSINR+~}
\def\discplnospace{DISC-MaxSINR+}

\newcommand{\R}{\mathbb{R}}
\newcommand{\C}{\mathbb{C}}

\newcommand{\figgref}[1]{Fig.~\ref{#1}}

\newcommand{\mR}{\mathbb{R}}

\SetKwInput{KwInput}{Input}
\SetKwInput{KwOutput}{Output}

\usepackage{xspace}
\usepackage[acronym,nogroupskip,nonumberlist,nopostdot]{glossaries}
\glsdisablehyper

\newacronym{awgn}{AWGN}{Additive White Gaussian Noise}
\newcommand{\awgn}{\gls{awgn}\xspace}
\newacronym{mmse}{MMSE}{Minimum Mean Square Estimate}

\newacronym{snr}{SNR}{Signal to Noise Ratio}
\newcommand{\snr}{\gls{snr}\xspace}
\newacronym{lts}{LTS}{Long Training Sequence}

\newacronym{sts}{STS}{Short Training Sequence}

\newacronym{fft}{FFT}{Fast Fourier Transform}

\newacronym{llr}{LLR}{Log Likelihood Ratio}

\newacronym{ofdm}{OFDM}{Orthogonal Frequency Domain Multiplexing}

\newacronym{rnn}{RNN}{Recurrent Neural Network}

\newacronym{bpsk}{BPSK}{Binary Phase Shift Keying}

\newacronym{ber}{BER}{Bit Error Rate}

\newacronym{sinr}{SINR}{Signal to Interference and Noise Ratio}
\newcommand{\sinr}{\gls{sinr}\xspace}
\newacronym{ml}{ML}{Maximum Likelihood}
\newcommand{\ml}{\gls{ml}\xspace}
\newacronym{mimo}{MIMO}{Multi-Input Multi-Output}

\newacronym{mdp}{MDP}{Markov Decision Process}
\newacronym{ne}{NE}{Nash Equillibrium}
\newacronym{mfe}{MFE}{Mean-Field Equillibrium}
\newacronym{mpe}{MPE}{Markov perfect equillibrium}
\newacronym{mfg}{MFG}{Mean-Field Game}
\newacronym{rl}{RL}{Reinforcement Learning}
\newacronym{marl}{MARL}{Multi-agent Reinforcement Learning}
\newacronym{iot}{IoT}{Internet of Things}
\newacronym{ssg}{SSG}{Stackelberg Security Game}
\newacronym{pse}{PSE}{Perfect Stackelberg Equilibrium}
\newacronym{mpse}{MPSE}{Markov \gls{pse}}
\newacronym{spbe}{SPBE}{Sub-game Perfect Bayesian Equilibrium}
\newacronym{pf}{PF}{Particle Filter}
\newacronym{pomdp}{POMDP}{Partially Obseravable \gls{mdp}}
\newacronym{nn}{NN}{Neural Network}
\newacronym{td}{TD}{Temporal Difference}
\newacronym{tim}{TIM}{Topological Interference Management}
\newacronym{gdof}{GDoF}{Generalized Degrees of Freedom}
\newacronym{dof}{DoF}{Degrees of Freedom}
\newacronym{inr}{INR}{Interference to Noise Ratio}
\newacronym{pam}{PAM}{Pulse Amplitude Modulation}
\newacronym{mse}{MSE}{Mean Square Error}
\newacronym{tdm}{TDM}{Time Domain Multiplexing}
\newacronym{gan}{GAN}{Generative Adversarial Network}
\newacronym{sir}{SIR}{Signal to Interference Ratio}
\newacronym{tin}{TIN}{Treating Interference as Noise}
\newacronym{ic}{IC}{Interference Channel}
\newacronym{ia}{IA}{Interference Alignment}

\usepackage{amsmath,amsfonts,bm}

\def\secref#1{Section~\ref{#1}}

\def\1{\bm{1}}

\def\mC{{\bm{C}}}

\def\mN{{\bm{N}}}

\def\mR{{\bm{R}}}

\DeclareMathAlphabet{\mathsfit}{\encodingdefault}{\sfdefault}{m}{sl}
\SetMathAlphabet{\mathsfit}{bold}{\encodingdefault}{\sfdefault}{bx}{n}

\def\sP{{\mathbb{P}}}

\newcommand{\E}{\mathbb{E}}

\usepackage{setspace}
\newcommand{\sjedit}[1]{{\color{black}#1}}

\begin{document}
\thispagestyle{empty}
\pagestyle{empty}
\title{Enhancing K-user Interference Alignment for Discrete Constellations via Learning}

\author{Rajesh Mishra, \sjedit{Syed Jafar}, Sriram Vishwanath, Hyeji Kim

\thanks{{Rajesh Mishra is with Samsung Research America Inc. Syed Jafar is with the Department of 
Electrical Engineering and Computer Science at the University of California at Irvine. Sriram Vishwanath and Hyeji Kim are with the Department of Electrical and Computer Engineering at the University of Texas at Austin, Austin, TX, USA. Emails: \{rajeshkmishra@utexas.edu, syed@uci.edu, sriram@utexas.edu, hyeji@utexas.edu\} } 
}

}

\maketitle
\thispagestyle{empty}

\begin{abstract}
    In this paper, we consider a $K$-user interference channel where interference among the users \sjedit{is neither too strong nor too weak}, a scenario that is relatively underexplored in the literature. We propose a novel deep learning-based approach to design the encoder and decoder functions that \sjedit{aim to} maximize the sumrate of the interference channel for discrete constellations. We first consider the \maxs algorithm, a state-of-the-art linear scheme for Gaussian \sjedit{inputs}, as the baseline and then propose a modified version of the algorithm for discrete \sjedit{inputs}. 
    We then propose a neural network-based approach that learns a constellation mapping \sjedit{with the objective of maximizing the sumrate. We provide numerical results to show that the constellations learned by the neural network-based approach 
    provide enhanced alignments, not just in beamforming directions but also in terms of the effective constellation at the receiver, thereby leading to improved sum-rate performance.}     
\end{abstract}

\begin{IEEEkeywords}
Multiuser interference channel, interference alignment, deep learning, neural networks, sumrate maximization.
\end{IEEEkeywords}
\section{Introduction}
\label{sec: Introduction}

With a renewed interest in multiuser and multi-cell scenarios such as MU-MIMO and Cell-Free Massive MIMO~\cite{elhoushy2021cell}, understanding the information-theoretic aspects of these systems has become increasingly important. Introduced in 1974~\cite{Ahlswede1974}, multiuser interference channels model multiple base stations concurrently interacting with numerous cellular mobiles. Extensive research has since been conducted to establish upper and lower capacity bounds for varying interference strengths~\cite{costa1985gaussian, sato1981capacity,etkin2008gaussian,shang2013two}. 

Most of these methods can be broadly divided into two categories: a) avoiding interference by \sjedit{projecting desired signal subspaces} orthogonal to the interfering signal space; b) managing interference by controlling the transmit power of the users and thus limiting the amount of interference seen at the receivers of the non-intended users. For $K$-user interference channels, interference alignment schemes were proposed~\cite{Cadambe2008a, Maddah-Ali2008, Cadambe2009} that tackle interference by projecting the signal onto interference-free subspaces, orthogonal to the interference from all the other interfering users in the channel. Distributed interference alignment schemes offer an iterative approach to derive the interference-free subspaces for transmission without requiring global channel knowledge~\cite{Gomadam2011}. Schemes provided by~\cite{Geng, Geng2013}, on the other hand, address interference through appropriate power allocation schemes.

Deep learning-aided physical layer (PHY) design has \sjedit{recently attracted much} interest for communication in general~\cite{nachmani2018deep,OShea-Erpek-Clancy2017,nachmani2019hyper,gan1,viterbinet,Elkelesh2019polar,kim2018deepcode,kim2018communication, wu2023cddm,du2023rethinking, liu2024cell} and for interference channels~\cite{OShea2017,Wu2020,zhang2023interference} in particular. A centralized autoencoder framework for two-user interference channels was introduced in~\cite{OShea2017}. They show that neural network-based codes, trained jointly, outperform time-sharing schemes for two-user interference channels. In~\cite{Wu2020}, the authors propose an adaptive deep learning algorithm for $K$-user symmetric interference channels and show that their algorithm outperforms the conventional system using $PSK$ or $QAM$. A joint optimization of the encoder-decoder setup in a deep autoencoder framework for a $2$-user Z-interference channel was proposed in~\cite{zhang2023interference}, where the authors find interference-aware constellations for finite-alphabet messages with imperfect channel knowledge. 

Optimal coding schemes for interference networks remain scarcely understood, especially in the moderate interference regime where the strength of the interfering signals is comparable to that of the desired signal from the users. In this paper, we primarily focus on such scenarios and aim to develop schemes for communicating discrete messages over interference channels. It is important to note that while most conventional work has focused on Gaussian inputs~\cite{mishra2021distributed}, we address more practical setups involving discrete inputs.

Starting with \maxs\cite{Gomadam2011}, the state-of-the-art distributed interference alignment scheme, we propose two modifications to enhance its achievable sumrate for discrete inputs. First, we employ a \ml decoder to improve the decoding efficiency. Second, we train a neural network that is initialized to mimic the \maxs algorithm to maximize the sumrate. The neural network learns suitable non-linear modulation maps, precoding vectors, and power allocations. We show that this method enhances the achievable sumrate and provides interpretations of the learned constellations. The rest of the paper is organized as follows. 

\begin{itemize}[leftmargin=*]

    \item{\em System model and problem formulation.} In Section~\ref{sec: System Model}, we describe the model for the $K$-user interference channel transmission system consisting of $K$ pairs of encoders and decoders. In Section~\ref{sec: problem formulation}, we formulate the problem where we describe the type of messages considered and introduce the encoder and decoder functions that are to be designed to optimize the objective function, which is maximizing the sumrate.

    \item {\em \maxs and \discnospace.} In Section~\ref{sec: background maxsinr}, we review the conventional interference alignment problem, where the messages are assumed to follow Gaussian distributions, and the state-of-the-art linear scheme, called the \maxs algorithm by~\cite{Gomadam2011}, which aims to optimize the Signal-to-Interference-and-Noise Ratio (SINR). While the Gaussianity assumption helps mathematical analysis, the messages often do not follow the Gaussian distribution. For example, the messages are often modulated symbols and can be modeled as uniform discrete random variables. 
    
    Considering these practical considerations, in Section~\ref{sec: discrete maxsinr}, we formulate a modified interference alignment problem, where the messages are assumed to follow discrete uniform distributions. As an initial attempt to solve this problem, we use the encoder designed using the \maxs algorithm as the baseline and then use the discrete messages as the input and \ml decoder for the decoding. This approach, we call \discnospace, forms the baseline for our future learning-based approach.

    \item {\em \discpl (\maxs with NN-based post processing)} In Section~\ref{sec: NN learning}, we ask the following question: ``can we go beyond the \maxs algorithm?" Specifically, we aim to design the encoder using a neural network framework with appropriate power control and solve the newly formulated optimization problem. We achieve this goal by (i) modeling the encoder as neural networks, (ii) presenting a modified analytical version of the \ml decoder, (iii) introducing a suitable loss function, and finally, (iv) training them jointly but with an initialization of the \maxs encoder and decoder.

    \item {\em Results and interpretation.} In Section~\ref{sec: Results}, we present the results of our numerical experiments and provide interpretations in Section~\ref{sec: Interpretation}.  
    We observe that the neural network-based approach leads to improved interference alignments, introduces varying sizes of constellations across users, and generates non-uniform constellations, resulting in a notable improvement in sumrate compared to the non-learning baselines. We investigate the roles of these three aspects and empirically demonstrate that each one contributes to the enhanced sum rate. Additionally, we show that incorporating insights from MaxSINR solutions as domain knowledge to initialize the neural networks is crucial for faster convergence and avoiding local minima.

\end{itemize}

\section{System Model}
\label{sec: System Model}
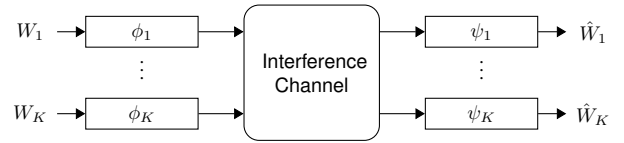
\begin{figure}[!htb]
    \centering
    \begin{tikzpicture}[scale = 0.75, transform shape,
    node distance = 2mm and 8mm,
         N/.style = {draw, minimum width=12mm, font=\sffamily, text width=5em, align=center},
       arr/.style = {draw, -{Triangle}}, 
       box/.style={rectangle, draw=none, text width=2em, text centered, minimum height=4em, minimum width=2em},
                            ] 
    \node (N)  [N,rounded corners=2mm, minimum size=24mm]    {Interference \\ Channel}; 
    \node (d)  [N, below left=of N.north west]               {$\phi_1$}; 
    \node (e)  [box,  left=.5cm of d]                        {$W_1$};
    \node (a)  [N, above left=of N.south west]               {$\phi_K$}; 
    \node (f)  [box,  left=.5cm of a]                        {$W_K$};
    \node (a3) [box] at ($(d)!0.4!(a)$) {\vdots};
    \node (b)  [N, above right=of N.south east] {$\psi_K$};
    \node (c)  [N, below right=of N.north east] {$\psi_1$};
    \node (a4) [box] at ($(b)!0.6!(c)$) {\vdots};
    \node (b1) [box, right=.5cm of b] {$\hat{W}_K$};
    \node (c1) [box, right=.5cm of c] {$\hat{W}_1$};
    \draw[arr]  (e) edge (e-|d.west)
                (f) edge (f-|a.west)
                (d) edge (d -| N.west) 
                (a) edge (a -| N.west) 
                (b -| N.east) edge (b)
                (c -| N.east) to (c);
    \draw[arr]  (b) edge (b1)
                (c) edge (c1);
        \end{tikzpicture} 
    \caption{Multiuser interference network setup with $m$-bit binary messages $W_i$ as inputs to the $K$ encoders ($\phi_{1:K}$) on the transmitter side which are received over an interference channel at the $K$ decoders ($\psi_{1:K}$) on the receiver side to produce the estimated messages $\hat{W}_i$.}
    \label{fig: System Setup}
\end{figure}

Consider a $K$-user interference channel depicted in~\figgref{fig: System Setup} with $K$ pairs of encoders and decoders at the transmitters and receivers, respectively. Let $\bld{Y}_i\in\C^n$ be the signal received at the $i^{\text{th}}$ decoder which can be expressed as
\begin{align}
    \bld{Y}_i = \sum_{j = 1}^K \bld{H}_{ij}\bld{X}_j + \bld{Z}_i \quad \forall i \in \sq{K}, \label{Eqn: receive equation for maxsinr}
\end{align}
where $\bld{X}_j\in\C^n$ denotes the transmitted symbol from the $j^{\text{th}}$ encoder, and $\bld{H}_{ij}\in\C^{n\times n}$ denotes the complex channel coefficient for the link between $i^{\text{th}}$ decoder and $j^{\text{th}}$ encoder. Let $\bld{Z}_i\in\C^n\sim \mC\mN \cq{0,\sigma^2\bld{I}}\ \forall i \in \sq{K}$ denote the Gaussian noise vector at $i^{\text{th}}$ receiver with variance $\sigma^2$ and $n$ be the symbol extension in the temporal or the spatial dimension. The power constraint on the transmitted symbol is given as $\E\sq{\|\bld{X}_j\|^2}\leq 1$, $\forall j \in [K]$.

We assume a global channel knowledge, i.e., the channel $\bld{H}_{ij}$, representing the link between the $i^\text{th}$ receiver and the $j^\text{th}$ transmitter for any $i$, $j$, is known at all the transmitters and the receivers. We also assume the channel to be reciprocal, as in~\cite{Gomadam2011}. Loosely speaking, if the receiver sees the least interference along a certain signaling dimension, it will also cause the least interference along that dimension when the roles of the transmitter and the receiver are switched. Mathematically, the received signal for a reciprocal channel can be expressed as
\begin{align*}
    \bld{Y}_j = \sum_{i = 1}^K \bld{\bar{H}}_{ji}\bld{X}_i + \bld{Z}_j \quad \forall j \in \sq{K},
\end{align*}
where the receivers switch roles as the transmitters, and $\bld{\bar{H}}_{ji}=\bld{H}_{ij}^T$, with $\cq{}^T$ denoting transpose. The reciprocity of channels is essential for distributed algorithms as it removes the need for the side information at the encoder about the amount of interference it causes to the unintended receivers. In the next section, we describe the assumptions on the message, encoder and decoder functions, and the performance metric, and we frame our objective.

\section{Problem formulation}
\label{sec: problem formulation}

We assume that the transmitted messages $W_i\in \{0,\cdots,2^{m}-1\}$ are $m$-bit binary messages. A suitable decoder is employed at the receiver to obtain the estimated message $\hat{W}_i$, which are also $m$-bit binary messages. Formally, let $ \phi_i$ denote the encoder and $\psi_i$ denote the decoder for the $i$-th user pair as shown in~\figgref{fig: System Setup}. The encoder $\phi_i$, defined as $\phi_i:W_i\in\bq{0,\cdots, 2^{m}-1}\to \bld{X}_i\in \C^n$, takes a discrete $m$-bit binary message sequence $W_i$ as input and maps it to a complex symbol $\bld{X}_i \in \C^n$ transmitted over the interference network. The decoder $\psi_i$, defined as $\psi_i:\bld{Y}_i\in\R^n\to \hat{W}_i\in\bq{0,\cdots, 2^{m}-1}$, maps the received signal $\bld{Y}_i\in\C^n$ to the estimated message bit sequence $\hat{W}_i$. Mathematically,
\begin{align}
    \bld{X}_i & = \phi_i\cq{W_i}, \label{Eqn: encoder equation}       \\
    \hat{W}_i & = \psi_i\cq{\bld{Y}_i}. \label{Eqn: decoder equation}
\end{align}

The encoder and decoder functions are designed to maximize a performance metric. This paper considers the {\em sumrate} as the performance metric and discusses how to design the encoder and decoder functions to maximize the sumrate.

\subsection{Sumrate as the performance metric and objective function}

We use the sumrate of the $K$-user interference channel both as the performance metric and the objective function to optimize the encoder and decoder functions. Specifically, we consider the sumrate that is given by the sum of the mutual information between the transmitted $m$-bit message $W_i$ and the estimated message bits $\hat{W}_i$, i.e.,
\begin{align}
    R :=\sum_{i=1}^K I\cq{W_i;\hat{W}_i}.\label{Eqn: sum rate equation}
\end{align}
From~\eqref{Eqn: sum rate equation}, we have the sumrate calculated as the sum of the mutual information between the transmitted and received messages for each user $i$. Mutual information is the maximum achievable rate for a given channel given a specific input distribution. In other words, it is the maximum number of bits that can be transmitted per channel use such that with an appropriate decoder, the error probability can be made arbitrarily small. In this paper, we assume that the input messages $W_i$'s are uniformly distributed, and we want to maximize the sumrate given such a distribution. The mutual information for each user $I\cq{W_i;\hat{W}_i}$ can be computed from the transition probability matrix, $\sP\cq{W_i,\hat{W}_i}\in \mR^{2^m}\times \mR^{2^m}$ as
\begin{align}
    I(W_i; \hat{W}_i) = \sum_{W\in\mathscr{W}}\sum_{\hat{W}\in\mathscr{W}} \sP(W,\hat{W})\log \frac{\sP(W,\hat{W})}{\sP(W)\sP(\hat{W})}, \label{Eqn: mutual information from probabilities}
\end{align}
where $\mathscr{W} = \{0,\cdots, 2^{m}-1\}$ is the set of possible discrete messages. The marginal probabilities, $\sP(W)$ and $\sP(\hat{W})$ are obtained by summing along the rows and columns of the transition probability matrix. Note that the joint probability matrix $\sP\cq{W_i\in\mathscr{W}, \hat{W}_i\in \mathscr{W}}$, with $\mathscr{W}=\{0,1,\cdots 2^{m}-1\}$, in~\eqref{Eqn: mutual information from probabilities} can be computed as
\begin{align}
    \sP\cq{W_i, \hat{W}_i} = \int \sP\cq{\hat{W}_i\vert \bld{Y}_i}\sP\cq{\bld{Y}_i\vert W_i}\sP\cq{W_i}d\bld{Y}_i, \label{Eqn: integration for mutual information}
\end{align}
where $\sP\cq{\bld{Y}_i\vert W_i}$ is the probability density function of the received symbols $\bld{Y}_i$ at the receiver $i$ given $W_i$ was the original message that was transmitted.
In the next section, we describe the \ml decoder, which minimizes the probability of decoding error in the
method to compute $\sP\cq{\bld{Y}_i\vert W_i}$.

\subsection{ML decoder as the optimal decoder for discrete messages}
\label{sec: ml decoder}
The $K$-user interference channel depicted in~\figgref{fig: System Setup} considers discrete messages as input. Before obtaining the optimal encoder that maximizes the sumrate in~\eqref{Eqn: sum rate equation}, we describe the \ml decoder $\psi_i = \psi_i^\text{ML}$ to recover discrete messages. The \ml decoding rule for estimating $m$-bit discrete messages $\hat{W}_i\in\{0,\cdots, 2^{m}-1\}$ for $i\in\sq{K}$ from the received signals $\bld{Y}_i=\bld{y}_i$ can be expressed as
\begin{align}
    \hat{W}_i\cq{\bld{y}_i} & = \underset{w^\prime_i\in \{0, \cdots, 2^m-1\}}{\arg \max}\sP\cq{\bld{Y}_i=\bld{y}_i\vert W_i=w_i^\prime}, \label{Eqn: Decoding rule for interference network}
\end{align}
where $\sP\cq{\bld{Y}_i=\bld{y}_i\vert W_i=w^\prime_i}$ denotes the probability of receiving the sample $\bld{Y}_i=\bld{y}_i$ given $W_i=w_i^\prime$ was the transmitted message while the the messages of other interfering users, $W_1, \cdots W_{i-1}, W_{i+1}, \cdots W_K$ are chosen at random. In the following, we describe the ML decoding rule and how to compute $\sP\cq{\bld{Y}_i\vert W_i}$.
\smallskip
\begin{theorem}
    \label{Thm: Decoding rule for K users}
    Let $W_i\in\bq{0,1,\cdots, 2^{m}-1}$ for $i \in \{1,\cdots,K\}$ be a set of discrete messages that are transmitted over a K-user interference channel with additive Gaussian noise. Let $\bld{Y}_i$ for $i \in \{1,\cdots,K\}$ denote the symbols that are received at the $i^{th}$ decoder, then the decision for the estimated message is taken based on the \ml decoding rule in~\eqref{Eqn: Decoding rule for interference network}, where the probability $\sP\cq{\bld{Y}_i\vert W_i}$ is computed as
    \begin{align}
        &\sP\cq{\bld{Y}_i\vert W_i=w_i} \propto \sum_{W_1=0}^{2^{m}-1}\cdots\sum_{W_{i-1}=0}^{2^{m}-1}\sum_{W_{i+1}=0}^{2^{m}-1}\cdots \nonumber\\ 
        &\sum_{W_{K}=0}^{2^{m}-1}  e^{-\frac{1}{2\sigma^2}||\bld{Y}_i - \bld{H}_{ii}\bld{V}_iW_i - \sum_{j = 1, i\neq j}^K \bld{H}_{ij}\bld{V}_jW_j||^2}. \label{Eqn: Pe distance formula}
    \end{align}
\end{theorem}

\begin{figure}[!htb]
    \centering
    \includegraphics[width=.2\textwidth]{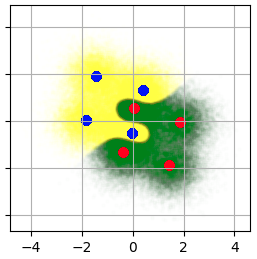}
    \caption{An illustration of non-linear ML decoding regions.}
    \label{fig: ml decoding decision region}
\end{figure}

\emph{Illustration:\ }Let us illustrate the decoding methodology through an example where three users transmit BPSK symbols over an interference channel. The encoders of each of these users encode a single-bit binary message into complex symbols, which are transmitted over the interference network and received at their respective decoders. The transmitted symbols interfere with one another, and each user receives a symbol different from the one transmitted. In~\figgref{fig: ml decoding decision region}, we plot the received symbols at the decoder of user $1$ when the transmitted message $W_1=0$ (red color) and $W_1=1$ (blue color) if there had been no noise. There are four dots of each color due to two interfering users causing four possible interfered symbols per each desired symbol; the interfering symbol pair $(W_2,W_3)$ could correspond to four different pairs $(0,0), (0,1), (1,0),$ or $(1,1)$. The colored patches represent all the possible received symbols when the transmitted message is $W_1=0$ and $W_1=1$, considering Gaussian noise.

An ideal decoder should have a decision boundary to correctly decode all the received interfered symbols to the transmitted message $W_1=0$ and $W_1=1$. The \ml decoder uses the \ml decoding rule, a distance-based metric, to compute the probability of each of those samples being from one of the transmitted messages $W_1=0$ or $W_1=1$. It considers the proximity of the received symbols to the desired set of symbols (blue and red dots in the figure) to compute the probability of each of the received symbols being from the transmitted message $W_1=0$ or $W_1=1$. The color codes, yellow and green, are done according to the probability $\sP\cq{\bld{Y}_1\vert W_1=0}$ or $\sP\cq{\bld{Y}_1\vert W_1=1}$ whichever is higher. In other words, the received samples that have a higher $\sP\cq{\bld{Y}_1\vert W_1=0}$ are coded yellow while the ones that have a higher $\sP\cq{\bld{Y}_1\vert W_1=1}$ are coded green. The decision boundary is the boundary between the yellow and the green patches. The final decision is made such that all received samples in the yellow region are mapped to $W_1=0$ while all the samples in the green region are mapped to $W_1=1$. This decoder strategy is followed throughout the paper and is used to optimize the objective function to compute an ideal encoder.

Now that we have an analytical description of the decoder, we are ready to formulate an objective function to obtain the optimal encoder that maximizes the sum rate in~\eqref{Eqn: sum rate equation}. The following section obtains the objective function for optimizing the encoder under the ML decoder.

\subsection{Objective}
\label{sec: objective}
Given the encoder $\phi_i$, and the decoder $\psi_i^\text{ML}$, for user $i$, we frame our final problem as
\begin{align}
    \phi_i & = \underset{\phi_i}{\arg \max} \sum_{i=1}^K I\cq{W_i;\psi_i^\text{ML}\cq{\bld{Y}_i}}, \label{Eqn: optimization problem}
\end{align}
where $\bld{Y}_i =\bld{H}_{ij}\bld{X}_i + \bld{Z}_i$, with $\bld{X}_i=\phi_i\cq{W_i}$, is the received signal at the $i^{th}$ decoder. The optimization problem in~\eqref{Eqn: optimization problem} is a non-convex optimization problem.

In the next section, we present a solution provided in~\cite{Gomadam2011} for the Gaussian messages and then propose modifications to the algorithm to make it suitable for discrete messages.

\section{MaxSINR: A state-of-the-art linear scheme for Gaussian messages}
\label{sec: background maxsinr}

In the previous section, we established our problem statement that we want to obtain an encoder that would maximize the sumrate of the $K$-user interference channel. In this section, we review existing approaches to interference alignment and establish baselines that we use later to evaluate our interference alignment algorithm. Specifically, in Section~\ref{sec: maxsinr algorithm}, we review the state-of-the-art interference algorithm, called the \maxs algorithm proposed in~\cite{Gomadam2011}; this algorithm aims to maximize the sum rate for Gaussian messages for such channels.
Unlike the setup in our paper, the \maxs algorithm assumes Gaussian messages and linear encoders and decoders. In Section~\ref{sec: discrete maxsinr}, we discuss how one can naturally tailor the \maxs algorithm to discrete messages by changing the decoder from linear to the maximum likelihood decoder for discrete messages, which we call Disc-MaxSINR.
This algorithm also becomes the baseline for our paper; in the following sections, we adapt the encoder and decoder to design an interference alignment algorithm tailored to the discrete messages and demonstrate the performance improvement upon Disc-MaxSINR.

\subsection{MaxSINR algorithm}\label{sec: maxsinr algorithm}

\noindent\textbf{Setup.\ } The paper~\cite{Gomadam2011} considers a $K$-user interference channel depicted in~\figgref{fig: System Setup} wherein the transmitter $i$ precodes a {\em Gaussian} message $W_i \sim \mathcal{N}(0,1)$ to a symbol $\bld{X}_i \in \C^n$ as $\bld{X}_i=\bld{V}_iW_i$ and at the receiver, the encoded message is retrieved from the received signal $\bld{Y}_i$ as $\hat{W}_i=\bld{U}_i\bld{Y}_i$.
The \maxs algorithm seeks to find the precoding vectors $\bld{V}_i\in \C^n$ and the combination vector $\bld{U}_i\in \C^n$ that maximize the sumrate of the interference channel assuming Gaussian transmitted symbols expressed as
\begin{align}
    R & = \sum_{i=1}^K I(\mathbf{X}_i;\mathbf{Y}_i),\label{Eqn: sum capacity equation maxsinr}
\end{align}
where $\bld{X}_i$ and $\bld{Y}_i$ are the transmitted and received symbols respectively. For the case when the transmitted and received symbols are Gaussian distributed, we have a closed-form expression for the mutual information of any user $i$ and can be rewritten as $I(\mathbf{X}_i;\mathbf{Y}_i) = \log \cq{1+\text{SINR}_i}$. Therefore, as the name of the algorithm suggests, the aim is to maximize the \sinr at each of the users such that the sum rate would be maximized. TABLE~\ref{tab: comparison of maxsinr setup and our setup} depicts the differences in our setup considered in the paper and the one in~\cite{Gomadam2011}.

\begin{table}
    \begin{center}
        \resizebox{0.5\textwidth}{!}{%
        \begin{tabular}{|l || l | l|}
            \hline
                               & \maxs                                   & Our Setup                         \\ [0.5ex]
            \hline\hline
            Messages           & $W_i \sim \mathcal{N}(0,1)$             & $W_i \in\{0,1,\cdots 2^{m}-1\}$   \\
            \hline
            Encoder            & $\bld{X}_i=\bld{V}_iW_i$                & $\bld{X}_i=\phi_i\cq{W_i}$        \\
            \hline
            Decoder            & $\hat{W}_i=\bld{U}_i\bld{Y}_i$          & $\hat{W}_i=\psi^{ML}_i\cq{Y_i}$   \\
            \hline
            Performance metric & $\sum_{i=1}^KI\cq{\bld{X}_i;\bld{Y}_i}$ & $\sum_{i=1}^KI\cq{W_i;\hat{W}_i}$ \\
            \hline
            \hline
        \end{tabular}
        }
    \end{center}
    \caption{Comparison of the \maxs setup and our setup}
    \label{tab: comparison of maxsinr setup and our setup}
\end{table}

\smallskip
\noindent\textbf{\maxs algorithm.\ } The \maxs algorithm proposed by authors in~\cite{Gomadam2011} aims to maximize the sum rate of the $K$-user network by finding the $\bld{U}_{i},\bld{V}_{i}$ vectors for all users, $i\in[1,K]$ that maximize the \sinr at each receiver $i$. The algorithm solves the optimization equation
\begin{align*}
    \bld{U}_{1:K}, \bld{V}_{1:K} = \underset{\bld{U}_{1:K}, \bld{V}_{1:K}}{\arg \max} \sum \log\cq{1+ \text{SINR}_i},
\end{align*}
where
\begin{align*}
    \text{SINR}_i  = \frac{\cq{\bld{U}_i\bld{H}_{ii}\bld{V}_i}\cq{\bld{U}_i\bld{H}_{ii}\bld{V}_i}^T}{\bld{U}_i\bld{B}_i\bld{U}_i^T + \sigma^2\bld{I}},
\end{align*}
and $\bld{B}_i = \sum_{j \neq i} \cq{\bld{H}_{ij}\bld{V_i}}\cq{\bld{H_{ij}V_i}}^T$ is the cumulative interference at the $i^{th}$ receiver.

\begin{algorithm}
    \SetAlgoLined
    \DontPrintSemicolon
    \For{$multiple$ $runs$}
    {
    Initialization: Choose $\bld{V}_{1:K}$ as $K$, $n\times 1$ random complex vectors\;
    \For{each iteration}
    {
    $\forall i\in [K]$, compute $\bld{U}_i$ as $\bld{U}_i = \frac{\bld{B}_i^{-1}\bld{H}_{ii}\bld{V}_i}{||\bld{B}_i^{-1}\bld{H}_{ii}\bld{V}_i||}$ \;
    Compute the sum rate: $R = \sum_{i=1}^K \log \cq{1 + \text{SINR}_i}$\;
    \textbf{Reciprocal channel:} $\forall i\in[K]$,  $\bld{\bar{V}}_i=\bld{U}_i$\;
    $\forall i\in [K]$, compute $\bld{\bar{U}}_i$ as $\bld{\bar{U}}_i = \frac{\bld{\bar{B}}_i^{-1}\bld{\bar{H}}_{ii}\bld{\bar{V}}_i}{||\bld{\bar{B}}_i^{-1}\bld{\bar{H}}_{ii}\bld{\bar{V}}_i||}$ \;
    \textbf{Forward channel:} $\bld{V}_i=\bld{\bar{U}}_i \ \forall i\in \sq{K}$\;
    }
    Store $\bld{V}_i$, $\bld{U}_i$ if $R$ is maximum
    }
    \caption{\maxs Algorithm}
    \label{alg: maxsinr algorithm}
\end{algorithm}

The computation of the precoding vectors $\bld{V}_i$ and the combination vectors $\bld{U}_i$ is done using an iterative algorithm and assuming reciprocity. Precisely, as depicted in Algorithm~\ref{alg: maxsinr algorithm}, the algorithm uses alternate maximization in the forward and reciprocal directions to ultimately converge to the final $\bq{\bld{V}}_{i=1}^{i=K}$ and $\bq{\bld{U}}_{i=1}^{i=K}$ which maximizes the sum-rate. The variables $\bar{\cq{\cdot}}$ denote the vectors computed for the reciprocal transmission.

Other interference alignment techniques work by suppressing the interference caused by any user to other users~\cite{Cadambe2008a, Schmidt2009,PetersSteven2009,Shi2011,Peters2011,Papailiopoulos2012}; however, they have poor performance at intermediate and low \snr regimes as no effort is made to maximize the desired signal strength at the receiver. \maxs maximizes the \sinr at the receivers of each of these users and, therefore, outperforms all these schemes in these regimes, though the optimality of \maxs algorithm is not guaranteed. Moreover, power allocation for users in this setting remains an open problem, which we will address later in our paper.

\subsection{\discnospace: Modifying \maxs for discrete messages}
\label{sec: discrete maxsinr}

In the previous section, we discussed the conventional \maxs algorithm, which is tailored for achieving the maximum sum rate for {\em Gaussian} messages. Here, we consider the setup introduced in~\secref{sec: System Model}, where the underlying messages are {\em $m$-bit discrete messages}, and introduce Disc-MaxSINR, a modified version of the \maxs algorithm. As we explain, we let the encoder be the MaxSINR encoder, while we use the ML decoder instead of the conventional MaxSINR decoder.

\begin{figure}[h]
    \centering
\begin{tikzpicture}[
  scale = 0.75, transform shape,
  shift={(0cm, 0cm)},  
       M/.style = {draw, minimum height = 5mm, minimum width=40mm, font=\sffamily, text width=15em, align=center},
       N/.style = {draw, minimum width=12mm, font=\sffamily, text width=5em, align=center},
     arr/.style = {draw, -{Triangle}}, 
     box/.style={rectangle, draw=none, text width=2em, text centered, minimum height=1em, minimum width=2em},
                          ] 
  \node (A)  [M, rotate = 90] at ($(-2,-2)$)   {Channel}; 
  \node (Z1) [circle, draw, inner sep=.5pt] at ($(A.north east)+(-0.65,-1)$) {+};
  \node (Z2) [circle, draw, inner sep=.5pt] at ($(A.north west)+(-0.65,+1)$) {+};
  \node (v1) [box] at ($(Z1.north) + (0,1)$)                                 {$\bld{V}_1$};
  \node (v2) [box] at ($(Z2.north) + (0,1)$)                                 {$\bld{V}_K$};
  \node (d1) [box] at ($(v2.north) + (0,1)$)                                 {$\vdots$};
  \node (B)  [M, rotate = 90] at ($(A.north) + (-1.5,0)$)                    {$m-$PAM}; 
  \node (w1) [box] at ($(B.north east) + (-1,-1)$)                           {$W_1$};
  \node (w2) [box] at ($(B.north west) + (-1,1)$)                            {$W_K$};
  \node (d3) [box] at ($(w2.north) + (0,1.8)$)                               {$\vdots$};
  \node (Y1) [circle, draw, inner sep=.5pt] at ($(A.south east)+(.65,-1)$)   {+};
  \node (Y2) [circle, draw, inner sep=.5pt] at ($(A.south west)+(.65,+1)$)   {+};
  \node (u1) [box] at ($(Y1.north) + (0,1)$)                                 {$\bld{U}_1$};
  \node (u2) [box] at ($(Y2.north) + (0,1)$)                                 {$\bld{U}_K$};
  \node (d2) [box] at ($(u2.north) + (0,1)$)                                 {$\vdots$};
  \node (C)  [M, rotate = 90] at ($(A.south) + (1.5,0)$)                     {ML Decoder}; 
  \node (w3) [box] at ($(C.south) + (1,0)$)                                 {$\sum_{i=1}^KI\left(W_i;\hat{W}_i\right)$};
  
  \draw[arr]  (w1) edge (w1-|B.north)
              (w2) edge (w2-|B.north)
              (Z1-|B.south) edge (Z1) 
              (Z2-|B.south) to (Z2)
              (Z1) edge (Z1-|A.north)
              (Z2) edge (Z2-|A.north)
              (Y1-|A.south) edge (Y1)
              (Y2-|A.south) to (Y2);
  \draw[arr]  (w3-|C.south) edge (w3)
              (Y1) edge (Y1-|C.north)
              (Y2) to (Y2-|C.north);
              \draw[arr]  (v1) edge (Z1)
              (v2) edge (Z2);
              \draw[arr]  (u1) edge (Y1)
              (u2) edge (Y2);
      \end{tikzpicture}
    \caption{\textit{\disc setup for discrete messages:} Beamforming vectors $\bq{\bld{V}_i}_{i=1}^K$ and $\bq{\bld{U}_i}_{i=1}^K$ are derived from the \maxs algorithm.}
    \label{fig: DiscMaxSINR network architecture}
\end{figure}
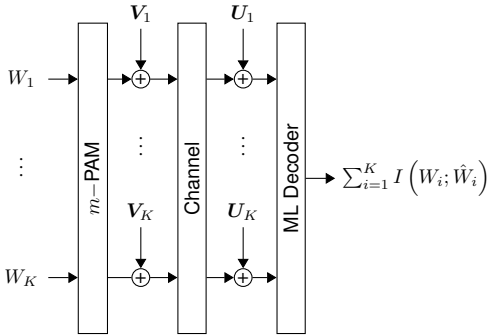

\noindent\textbf{Setup.\ }The \disc setup for the discrete messages is shown in~\figgref{fig: DiscMaxSINR network architecture} where we show the Gaussian interference channel with $K$ pairs of encoders and decoders. The equation governing the received signal $\bld{Y}_i$ at decoder $i$ is the same as in~\eqref{Eqn: receive equation for maxsinr}. We focus on communicating $m$-bit binary messages, i.e., $W_i \in \{0,\cdots, 2^{m}-1\}$ for all $i \in [K]$.

\vspace{.2em}
\smallskip
\noindent\textbf{Encoder as the MaxSINR encoder.\ }
The encoder first modulates the $m$-bit binary messages, i.e., $W_i \in \{0,\cdots, 2^{m}-1\}$ for all $i \in [K]$ using the $m$-PAM and then precodes it with the $\bld{V}_i$'s obtained from the algorithm described in~\secref{sec: background maxsinr}. The $i^{th}$ encoder of the \disc algorithm is given as {$\bld{X}_i=\sqrt{P}_i\bld{V}_iW_i^\star$}, where the precoding vectors $\{\bld{V}_i\}_{i=1}^K$, $\bld{V}_i\in\R^n$ are derived from the conventional \maxs algorithm ($n$ is the symbol extension) and $W_i^\star\in\R$ are the PAM-modulated symbols of the discrete $m-$bit binary messages $W_i$, normalized to unit power.

\vspace{.2em}

\smallskip
\noindent\textbf{Decoder as the ML decoder.\ }
The design of the decoder for the \maxs setup can be done in two ways. We can assume a {\em linear receiver} where the received signal $\bld{Y}_i$ is decoded using the combination vector $\bld{U}_i$ as $\hat{W}_i=\bld{U}_i\bld{Y}_i$ which is then sent to \ml decoder that obtains the estimated message $\hat{W}_i$. The other way is to assume a {\em non-linear ML receiver} where the received signal $\bld{Y}_i$ is sent to the \ml decoder directly to obtain the estimated message $\hat{W}_i$. The benefit of the ML decoder can be seen in the illustration in~\secref{sec: ml decoder}. We consider both cases and compare the performance of the two receivers in the next section. We, henceforth, refer to the linear receiver as \maxs while the non-linear receiver representation of the \maxs setup is referred to as \disc.

In summary, the \disc setup provides a conventional encoder design algorithm using the \maxs algorithm but using the \ml decoder and the discrete message setup considered in the paper. In the subsequent sections, we focus on optimizing the encoder design for the discrete messages. We first propose a neural network-based approach to learn the encoder constellation for discrete message transmission.

\section{Disc-MaxSINR+: Enhancing \maxs with Neural Network Learning}
\label{sec: NN learning}

In the previous section, we proposed modifications to the \maxs algorithm to make it suitable for discrete messages. However, like the conventional algorithm, it still adheres to a linear encoder mapping between the messages and the transmitted symbols.

To further improve the performance of communicating discrete messages over the $K$-user interference channel, we relax the linearity constraint on the encoder and propose a neural network-based approach to learn the constellation jointly with a suitable precoder and power allocation. We term this approach as ``\discplnospace''. We first describe the neural network architecture consisting of the encoder and the decoder. We then describe the training procedure and the loss function. We base our neural network architecture on the encoder architecture of the \maxs setup and use the \ml decoder described in~\secref{sec: ml decoder}.

\begin{figure*}[!ht]
    \centering
    \def\svgwidth{.7\textwidth}
    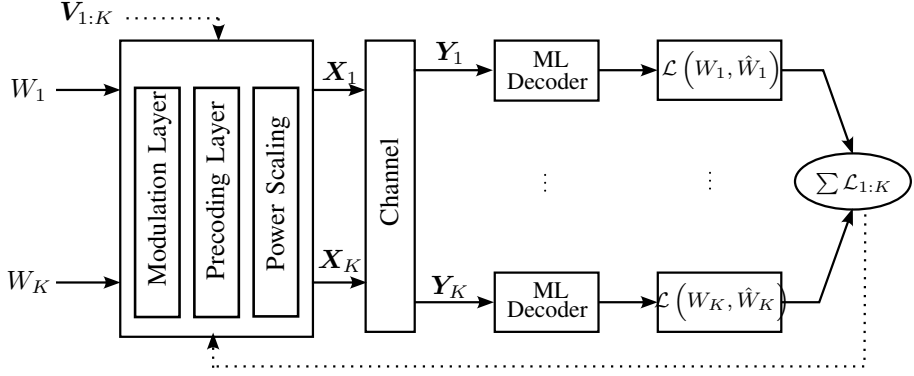
    \caption{\discpl: Neural Network Architecture. It consists of the encoder part, which is a three-layer neural network, and the decoder part, which is the \ml decoder. The encoder part consists of a modulation layer, a precoding layer, and a power allocation layer. The loss function is computed for each user, and the sum of the loss functions is minimized.}
    \label{fig: Discmaxsinr_plus Encoder Architecture}
\end{figure*}

\subsection{Encoder}
The neural network architecture, as shown in~\figgref{fig: Discmaxsinr_plus Encoder Architecture}, is broadly divided into three parts: the encoder, the decoder, and the computation of loss function. 
The encoder operation is represented as
\begin{align*}
    \bld{X}_i = \phi_i\cq{W_i; \bld{M}_i \bld{V}_i, P_i},
\end{align*}
where $W_i$ is the input symbol to be transmitted, $\bld{M}_i\in\mR^{2^m}$, $\bld{V}_i\in\mR^n$ and $P_i\in\mR$ are the parameters of the encoder. It can be viewed as a three-layer neural network as shown in~\figgref{fig: Discmaxsinr_plus Encoder Architecture} consisting of a modulation layer, a precoding layer, and a final power allocation layer. The modulation layer maps the input message $W_i$ to one of the $2^{m}$ constellation points on the real line. The precoding layer precodes it with a $n$ dimensional vector, and the final power allocation layer scales the transmit vectors using the power allocation parameters $P_i$. As we explain in detail later in Section~\ref{sec: training}, the architecture takes the learned precoders and the $m-$PAM constellation as the pretraining inputs for setting the initial conditions for the trainable parameters. We now describe the mathematical operations involved in each of these layers, the parameters that need to be learned, and how some of the parameters are initialized.

\textbf{Modulation layer:} The input to the Modulation layer is $W_i^\dag\in\mR^{2^m}$, which denotes a one hot representation of the input message $W_i$. That is, $W_i = m$ corresponds to $W_i^\dag$ such that all elements are zero except the $m^\text{th}$ element which is set to 1. Let $\bld{M}_i\in\mR^{2^m\times 1}$ be a single-dimensional vector containing the learned constellation. The output of the modulation layer is given by $\tilde{W}_i = \bld{M}_i^TW_i^\dag$ where $\tilde{W}_i\in \mR$. The parameters of $\bld{M}_i$ need to be learned with the only constraint that $\E\cq{\tilde{W}_i} = 0$. In order to ensure that, we enforce the constraint $\{m_i^{2^{m-1}+1}:m_i^{2m}\} = -\{m_i^0:m_i^{2^{m-1}}\}$ where $m_i^j$ represents the $j^{\text{th}}$ element of $\bld{M}_i$. These parameters are initialized with the values of a $2^m$-PAM constellation. It is also ensured that the learned constellation points are normalized to unit power at each step of the training process.

\textbf{Precoding layer:} The input to the precoding layer is $\tilde{W}_i$ and the output is $\tilde{\bld{X}}_i\in\mR^n$ which is obtained as $\tilde{\bld{X}}_i = \bld{V}_i \tilde{W}_i$. The parameters of the precoding layer are $\bld{V}_i\in\mR^{n\times 1}$ which needs to be learned with the constraint that $\E\cq{\|\tilde{\bld{X}}_i\|^2} = 1$. This is ensured by normalizing the precoding vector $\bld{V}_i$ after each update. The parameters are initialized with the precoding vectors obtained from the \maxs algorithm.

 \textbf{Power scaling:} The power allocations are done across users to maximize the sum rate. The power allocation parameters are $P_i\in\mR$, which needs to be learned. These values are initialized with unity and then learned during the training process. The values are constrained to be between 0 and 1.

\subsection{Decoder}
For the decoder, we adopt a modified version of the \ml decoder described in Section~\ref{sec: ml decoder}. The decoder with the decoding rule in~\eqref{Eqn: Decoding rule for interference network} involves an argmax operation that renders the training process non-differentiable. Therefore, we use an \ml decoder with ``soft decoding" that outputs a probability distribution over the possible transmitted messages. We then apply a temperature scaling to the probability distribution to mimic the argmax operation. The temperature scaling is replaced by the argmax operation in the final testing scenarios to compute the sumrate.

Let us define a function $\bld{Q}_i:\mR^n \rightarrow \mR^{2^m}$ that maps the received sample $\bld{Y}_i=\bld{y}_i$ to a scaled probability distribution over the possible transmitted messages.
Letting $\bld{Q}_i^{(l)}(\bld{y}_i)$ denote the the $l$-th element of $\bld{Q}_i(\bld{y}_i)$, i.e., $\bld{Q}_i(\bld{y}_i) = (\bld{Q}_i^{(0)}(\bld{y}_i), \cdots, \bld{Q}_i^{(2^m-1)}(\bld{y}_i))$, we let
\begin{align}
    \bld{Q}_i^{(l)}\cq{\bld{y}_i} &= \frac{e^{-\beta_i \sP\cq{\hat{W}_i = l \vert \bld{Y}_i=\bld{y}_i}}}{\sum_{l=0}^{2^m-1} e^{-\beta_i \sP\cq{\hat{W}_i = l \vert \bld{Y}_i=\bld{y}_i}}} \label{Eqn: transition probability equation as output of decoder}
\end{align}

\noindent for $l = 0, \cdots, 2^m-1$. The parameter $\beta_i$ is a temperature scaling parameter that helps us to mimic the argmax operation without making the process non-differentiable. As the value, $\beta_i$ increases, the probability distribution becomes more peaked around the maximum value which converges to the argmax operation as $\beta_i\rightarrow \infty$. With $\beta_i$ set to zero, the probability distribution becomes uniform. This way, we relax the argmax operation to a varying degree, where the amount of variation can be learned during the training process. In the final testing scenarios, the temperature scaling is replaced by the argmax operation to compute the sum rate.

The computed probabilities are computed analytically using the sum rate via the mutual information in~\eqref{Eqn: sum rate equation}. By relaxing the argmax operation during the training, we ensure that the whole process is differentiable so that the gradient can be computed and the parameters can be updated using backpropagation. The following section describes the loss function and the training process.

\subsection{Loss function}
As our goal is to maximize the sum rate, defined in~\eqref{Eqn: sum rate equation}, we define the loss function as the negative sum rate. That is, we let
\begin{align*}
    \mathcal{L}_i &= -I\cq{W;\hat{W}_i} \\
    &= -\sum_{w_i,w_i^\prime} \sP\cq{W_i=w_i,\hat{W}_i=w_i^\prime}\\
    &\ \ \ \ \ \ \times \log\cq{\frac{\sP\cq{W_i=w_i, \hat{W}_i=w_i^\prime}}{\sP\cq{W_i=w_i}\sP\cq{\hat{W}_i=w_i^\prime}}}
\end{align*}
and our final loss function is given as $\mathcal{L}=\sum_i\mathcal{L}_i.$

The key term in evaluating the loss function is the transition probability matrix, $\sP\cq{W_i, \hat{W}_i}\in \mR^{2^m}\times \mR^{2^m}$, which refers to the joint probabilities of transmitting $W_i=w_i$ and estimating it as $\hat{W}_i=w_i^\prime$ for all possible values of $w_i$ and $w_i^\prime$.

Note that each element of the probability transition matrix can be written in terms of the decoding probabilities $\bld{Q}_i$ as
\begin{align}
    &\sP\cq{W_i=w_i, \hat{W}_i=l}  \nonumber\\
    &= \int \bld{Q}_i^{(l)}\cq{\bld{y}_i} \sP(\bld{Y}_i = \bld{y}_i|W_i = w_i)\sP\cq{W_i=w_i}d\bld{y}_i. \label{Eqn: integration prob transition equation}
\end{align}

We can then approximate~\eqref{Eqn: integration prob transition equation} from samples via Monte Carlo simulation as
\begin{align}
    &\sP\cq{W_i=w_i, \hat{W}_i=l} 
     \approx \frac{1}{N} \sum_j \bld{Q}_i^{(l)}\cq{\bld{y}_i^j}\sP\cq{W_i=w_i}, \label{Eqn: summation prob transition equation 2}
\end{align}
where $\bld{y}_i^j$ represents the $j^\text{th}$ sample received at the receiver $i$, $j\in\sq{N}$, and $N$ is the number of sample points used to compute the sample average. We can also calculate an entire row of the transition probability matrix simultaneously: noting $\bld{Q}_i(\bld{y}_i) = (\bld{Q}_i^{(0)}(\bld{y}_i), \cdots, \bld{Q}_i^{(2^m-1)}(\bld{y}_i))$,

\begin{align}
   & \left\{\sP\cq{W_i=w_i, \hat{W}_i=l}\right\}_{ l \in [0,2^m-1]} \nonumber\\
    & \approx \frac{1}{N} \sum_j \bld{Q}_i\cq{\bld{y}_i^j}\sP\cq{W_i=w_i}. \label{Eqn: summation prob transition equation 3}
\end{align}

This is a Monte Carlo approximation of the transition probability matrix and represents a single row of the matrix corresponding to the transmitted message $W_i=w_i$. It is repeated for each possible transmit message to complete the matrix.
At the end of this operation, we have a matrix where the rows correspond to the transmitted symbols and the columns correspond to the received symbols. The matrix is populated with the probabilities of receiving the symbol $\hat{W}_i = l $ given that the transmitted symbol was $W_i=w_i$.

\subsection{Training}
\label{sec: training}

We train the network using a batch size of $10000$ samples of $2^m \times K$ one hot vectors as input to the encoder. These are obtained by generating a random set of binary messages $W_i\in \bq{0,1}^{m}$ and then setting the corresponding position of a zero vector as $1$. Each epoch consists of $2$ steps. The first step is to send a set of message bits that consist of all the combinations of binary messages across all the users so that the reference points are captured at the decoder. In the next step, we send the input samples and obtain the loss function. An Adam optimizer with a fixed learning rate of $0.001$ is used to compute the gradient and update the encoder's parameters.

\noindent \textbf{Pretraining.} An important aspect of the training is that we begin the neural network training by loading the network parameters with a set of constellation points obtained by the \disc algorithm instead of random initialization. We term the step as ``pretraining". The training of the neural network representing the encoder has trainable variables corresponding to the modulation layer, which is the mapping of the messages to modulated symbols; the precoding layer, where the symbols are precoded with the precoding vectors, and the power allocation layer, where the symbols are scaled with the power allocation parameters. The naive way of initializing them before training is to assume random values for these variables. However, we find that the training process in such a case is slow, and there is a high chance that the network will get stuck in a local minima. Therefore, we initialize the trainable variables with the constellation points obtained from the \disc algorithm, where the variables in the precoding layers are initialized with the precoding vectors obtained from the \disc algorithm. In contrast, the variables in the modulation layer are initialized with the constellation points of the corresponding PAM constellation. This ensures that the network does not start with a random constellation, making the training faster. Moreover, this also ensures that the optimization is not stuck in local minima. We empirically verify that this pretraining phase is crucial.

\section{Results}
\label{sec: Results}
In this section, we present the results of our work. We showcase the plots of sumrate using the approaches discussed in Section~\ref{sec: background maxsinr} and Section~\ref{sec: NN learning}. The comparison is performed for two different channel settings, the symmetric and asymmetric interference channels, which we describe in Section~\ref{sec: channel characterization}. In Section~\ref{sec: channel characterization}, we describe the channel we consider, the effect of different channel parameters on the channel performance, and ways to characterize them. In Section~\ref{sec: performance evaluation}, we show a superior performance of our proposed network-based approaches over the conventional \maxs algorithm; we show that choosing a non-linear mapping for the messages in the constellation space leads to a noticeably better performance in terms of sum rate.

The plots compare with the baseline technique, which uses a distributed iterative approach to obtain the precoding vectors that maximize the \sinr at the receiver. It is the conventional \maxs algorithm proposed in~\cite{Gomadam2011}. Though the computation of the precoding vectors is independent of the type of input symbols, the algorithm is optimal for Gaussian symbols. We compare this approach to study the performance difference in the sumrate if we use a non-linear receiver instead of a linear receiver or a non-linear mapping of the transmitted symbols to the constellation space.

\subsection{Channel characterization}
\label{sec: channel characterization}

In this paper, we have mainly focused on the \awgn channels, where each received signal $\bld{Y}_i$ is expressed as
\begin{align}
    \bld{Y}_i = \sum_{j = 1}^K \bld{H}_{ij}\bld{X}_j + \bld{Z}_i \quad \forall i \in \sq{K}, \label{Eqn: received signal equation}
\end{align}
where the strength of the channel, $\bld{H}_{ij}\in\mC^{n\times n}$ is the complex channel coefficient for the link between receiver $i$ and transmitter $j$, $n$ being the symbol extension in the spatial or temporal dimension, while $\bld{Z}_i\in\C^n\sim \mC N\cq{0,\sigma^2\bld{I}}\ \forall i \in \sq{K}$ is the Gaussian noise vector at receiver $i$. The power constraint on the transmitted symbol is given as $\E\sq{||\bld{X}_i||^2}\leq 1$.

The channel coefficient $\bld{H}_{ij}$ of link $ij$ is represented as $\bld{H}_{ij} = \bld{R}_{ij}e^{\bld{\theta}_{ij}}$ where $\bld{R}_{ij}$ and $\bld{\theta}_{ij}$ are the channel gain and phase parameters respectively. Let us now describe these parameters in terms of the channel \snr, the strength of the interfering signal on the crosslinks ($\bld{\alpha}_{ij}$), and the phase difference induced by the channel.
\begin{figure}
    \centering
    \begin{tikzpicture}[scale = .8, transform shape]
        \node(img) at (0,0) { \includegraphics[width=.45\textwidth]{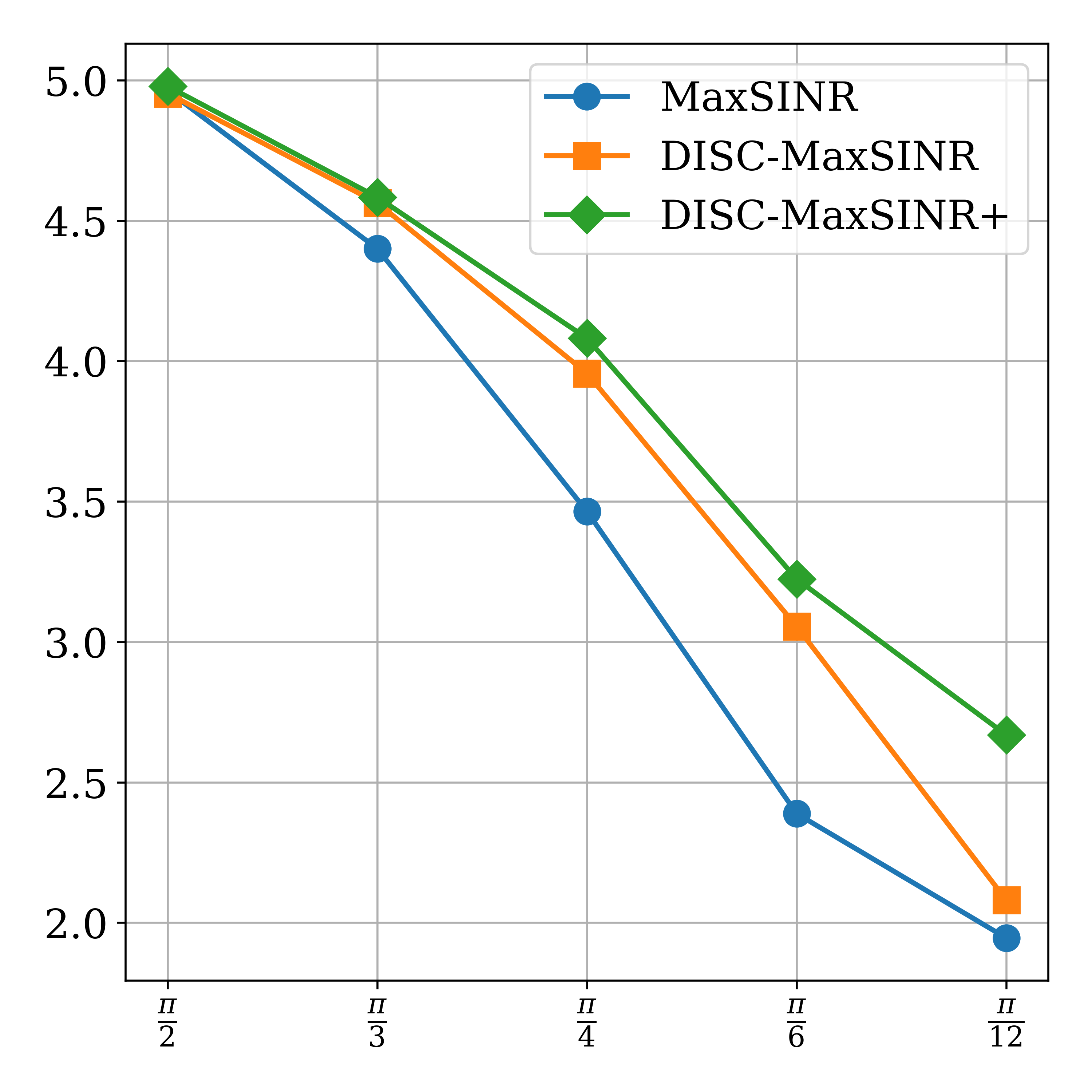}};
        \node[below=of img, node distance=0cm, yshift=1.4cm] {Channel rotation, $\theta$};
        \node[left=of img, node distance=0cm, rotate=90, anchor=center,yshift=-1.1cm]{Sumrate (bits/s/Hz)};
        \node[above=of img, node distance=10cm, yshift=-1.4cm] {$\text{SNR}=12 \text{dB}$, $4-$PAM, $3$ users};
    \end{tikzpicture}
    \begin{tikzpicture}[scale = .8, transform shape]
        \node(img) at (5,0) { \includegraphics[width=.45\textwidth]{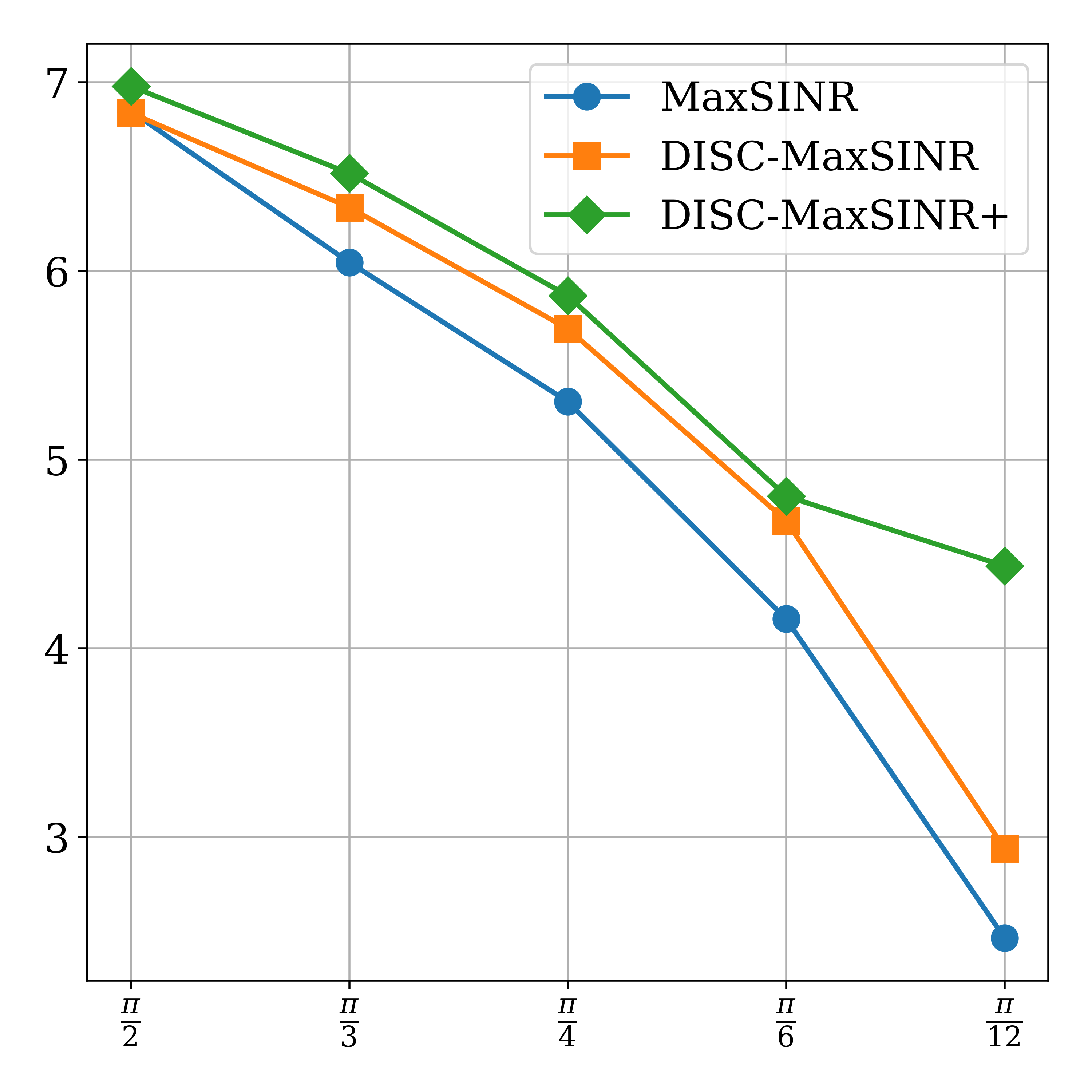}};
        \node[below=of img, node distance=0cm, yshift=1.4cm] {Channel rotation, $\theta$};
        \node[left=of img, node distance=0cm, rotate=90, anchor=center,yshift=-1.1cm]{Sumrate (bits/s/Hz)};
        \node[above=of img, node distance=10cm, yshift=-1.4cm] {$\text{SNR}=16 \text{dB}$, $8-$PAM, $3$ users};
    \end{tikzpicture}
    \caption{Comparison of sumrate for $K=3$ users between \maxs (discrete messages with linear receiver), \disc (discrete messages with \ml decoder) and \discpl (with neural encoders and power allocation) when $m= 4, 8$ for different SNRs with symmetric channel settings. It can be seen that the \disc and \discpl algorithms improve performance compared to conventional algorithms.}
    \label{fig: NN comparison with maxsinr with symmetric channels}
\end{figure}
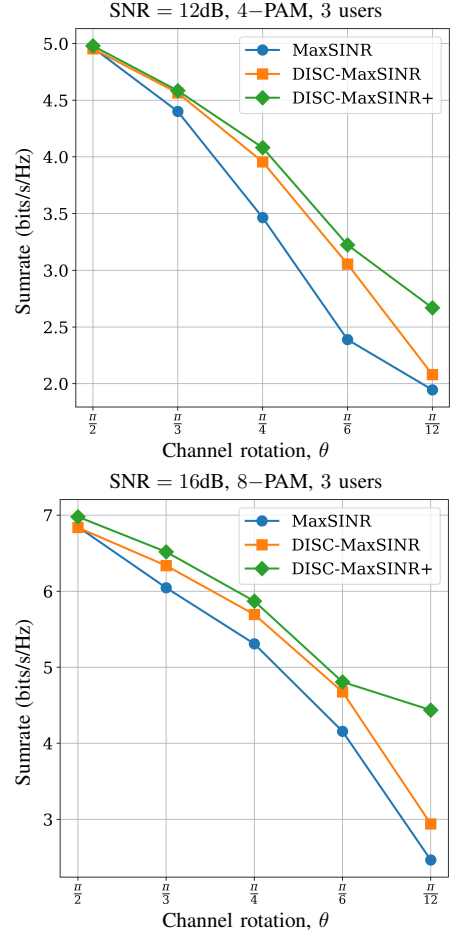

We define the channel \snr as the amount of the additive noise in the direct links. We assume the direct channel links $\bld{H}_{ii}$ have a unity channel gain i.e. $R_{ii}=\bld{1}_n \ \forall i\in\sq{K}$ and no phase i.e. $\bld{\theta}_{ii}=0 \ \forall i\in\sq{K}$ where $\bld{1}_n$ is a $n\times n$ matrix of ones. We assume all the direct links for all the encoder-decoder pairs have the same channel \snr. Therefore, we can express the noise variance as

\begin{align}
    \text{SNR} & = \frac{||\bld{R}_{ii}||^2}{\sigma^2}  := \frac{1}{\sigma^2}, \forall i\in\sq{K}. \label{Eqn: Channel SNR from sigma}
\end{align}

Next, we define the interference strength parameter $\bld{\alpha}_{ij}$ as
\begin{align*}
    \bld{\alpha}_{ij} = \frac{\log \bld{\text{INR}}_{ij}}{\log \text{SNR}},
\end{align*}
where $\bld{\text{INR}}_{ij}$ can be expressed as $\bld{\text{INR}}_{ij} = \frac{||\bld{R}_{ij}||^2}{\sigma^2}$. It signifies the relative strength of the signal in the interfering links relative to the direct link. The value of the parameters in $\bld{\alpha}_{ij}$ varies from $\infty$ to $-\infty$. A value of $1$ signifies that the strength of the interfering signal from link $ij$ is the same as the desired link $i$. We can determine the channel gain parameters of the cross-links $\bld{R}_{ij}$, $i\neq j$ in terms of $\bld{\alpha}_{ij}$ in an element-wise operation given as
\begin{align}
    \bld{R}_{ij} = \bld{R}_{ii} 10^{0.5\log \text{SNR}\cq{1-\bld{\alpha}_{ij}}}. \label{Eqn: channel from alpha}
\end{align}
Given the information about $\bld{\alpha}_{ij}$, $\bld{\theta}_{ij}$ and $\sigma^2$, we can determine the channel coefficients using~\eqref{Eqn: Channel SNR from sigma} and~\eqref{Eqn: channel from alpha}.

\subsection{Performance Evaluation}
\label{sec: performance evaluation}

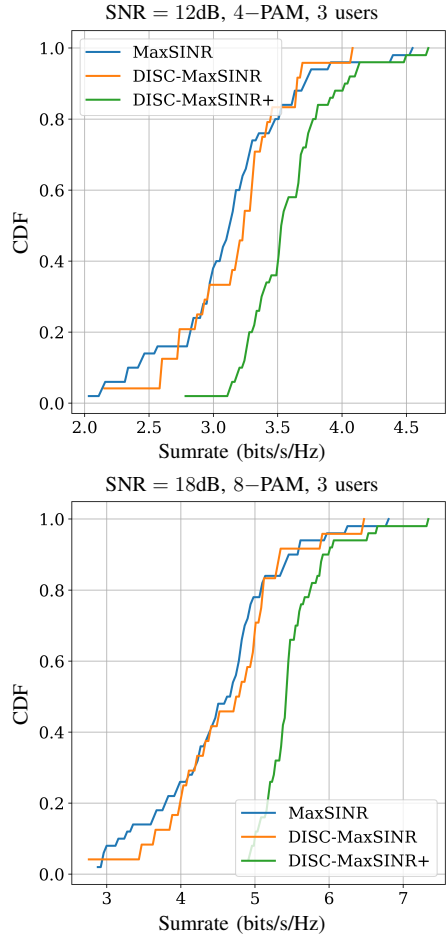
\begin{figure}[!htb]
    \centering
    \begin{tikzpicture}[scale = .8, transform shape]
        \node(img) at (0,0) { \includegraphics[width=.45\textwidth]{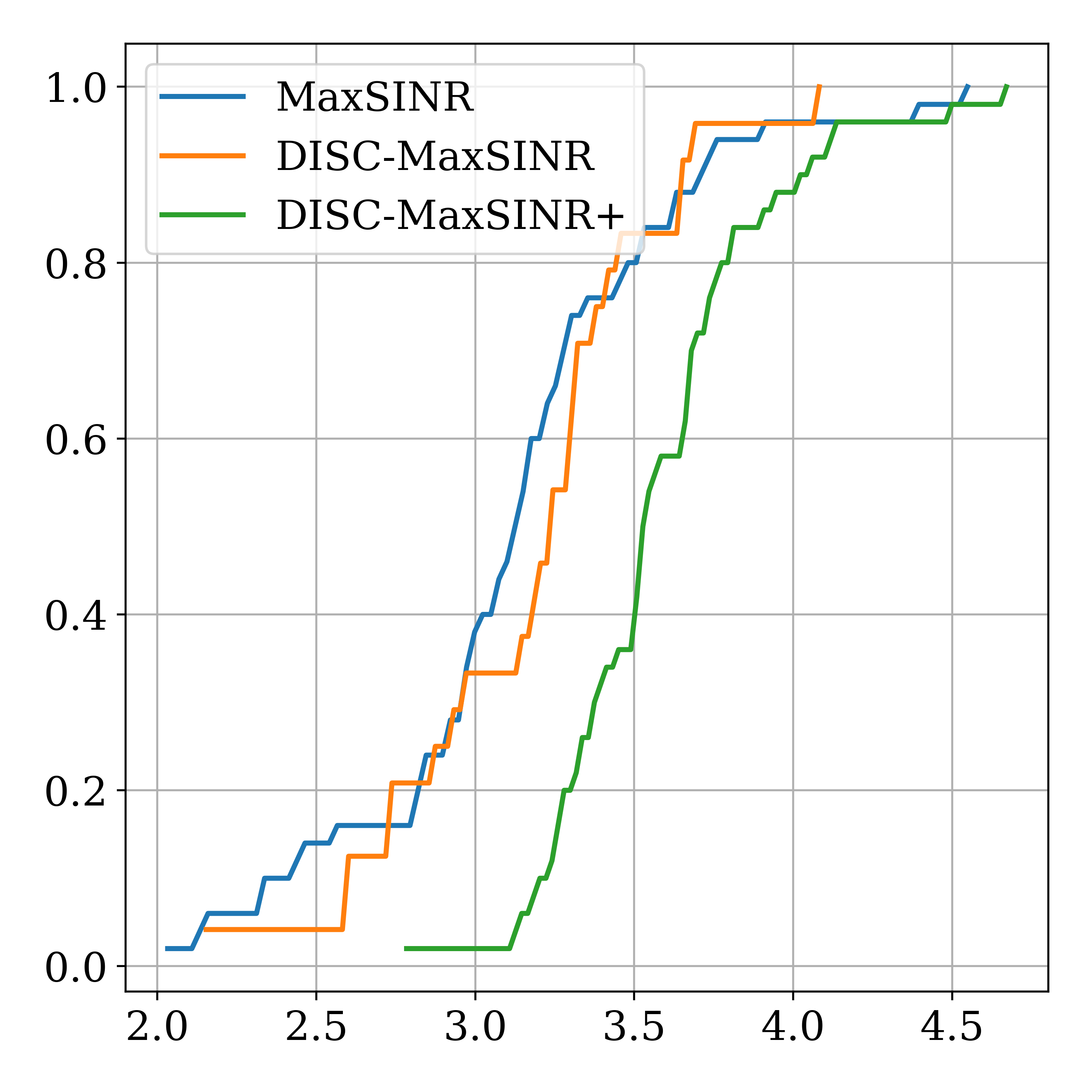}};
        \node[below=of img, node distance=0cm, yshift=1.4cm] {Sumrate (bits/s/Hz)};
        \node[left=of img, node distance=0cm, rotate=90, anchor=center,yshift=-1.1cm]{CDF};
        \node[above=of img, node distance=10cm, yshift=-1.4cm] {$\text{SNR}=12 \text{dB}$, $4-$PAM, $3$ users};
    \end{tikzpicture}
    \begin{tikzpicture}[scale = .8, transform shape]
        \node(img) at (5,0) { \includegraphics[width=.45\textwidth]{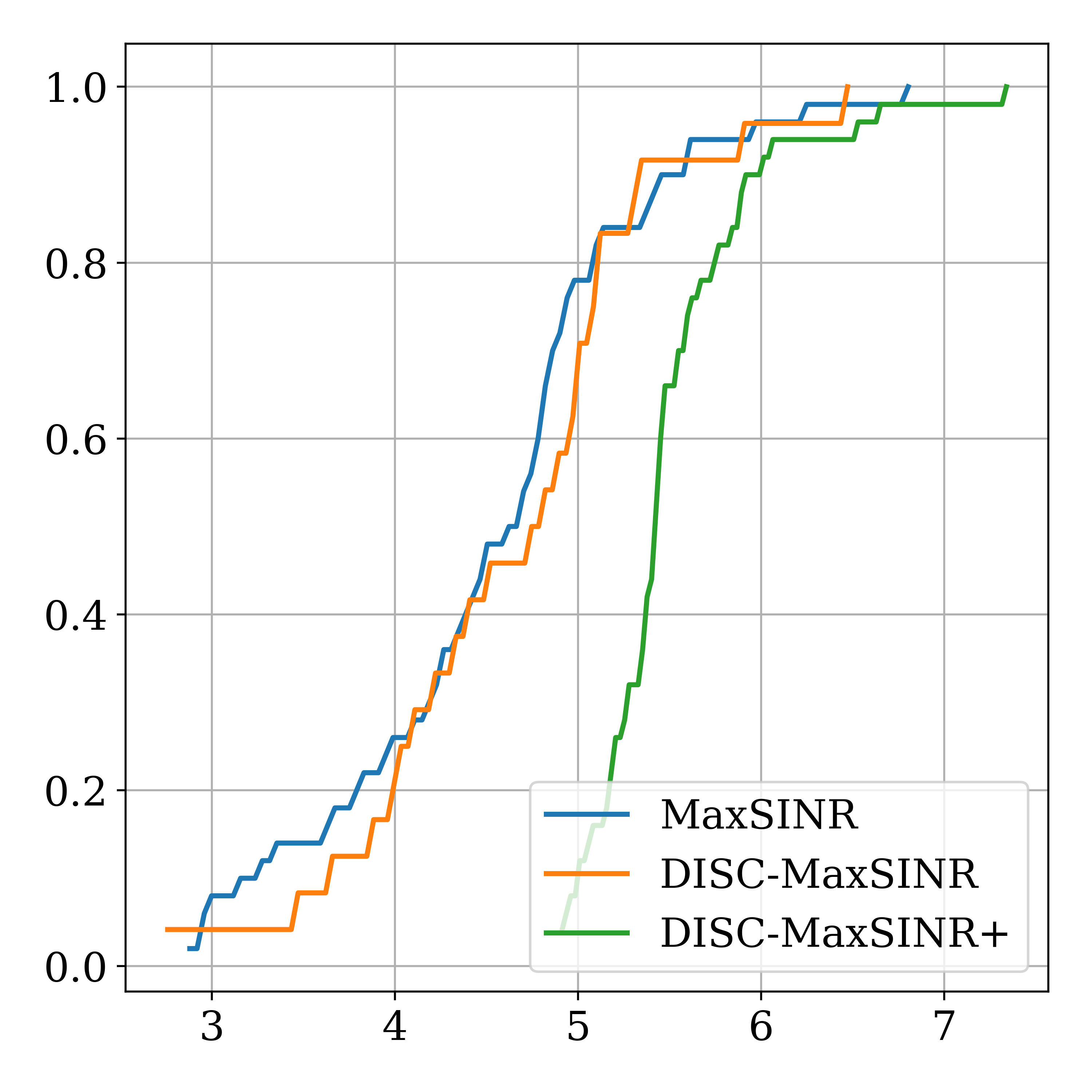}};
        \node[below=of img, node distance=0cm, yshift=1.4cm] {Sumrate (bits/s/Hz)};
        \node[left=of img, node distance=0cm, rotate=90, anchor=center,yshift=-1.1cm]{CDF};
        \node[above=of img, node distance=10cm, yshift=-1.4cm] {$\text{SNR}=18 \text{dB}$, $8-$PAM, $3$ users};
    \end{tikzpicture}
    \caption{Comparison of sumrate for $K=3$ users between \maxsnospace, \discnospace, and \discpl when $m= 4, 8$ for different SNRs with asymmetric channel settings. The comparison shows that retraining the precoding vectors obtained from \maxs greatly improves the sumrate performance.}
    \label{fig: NN comparison with maxsinr with asymmetric channels}
\end{figure}

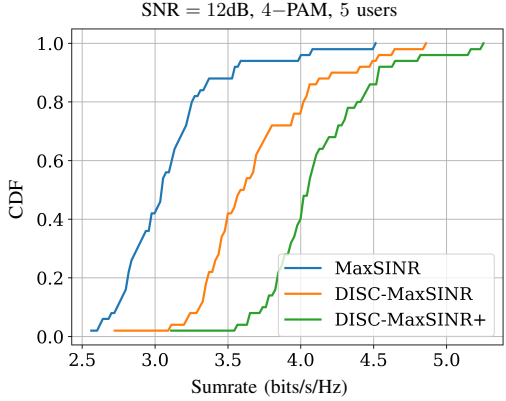
\begin{figure}[!htb]
    \centering
    \begin{tikzpicture}[scale = .75, transform shape]
        \node(img){ \includegraphics[width=.55\textwidth]{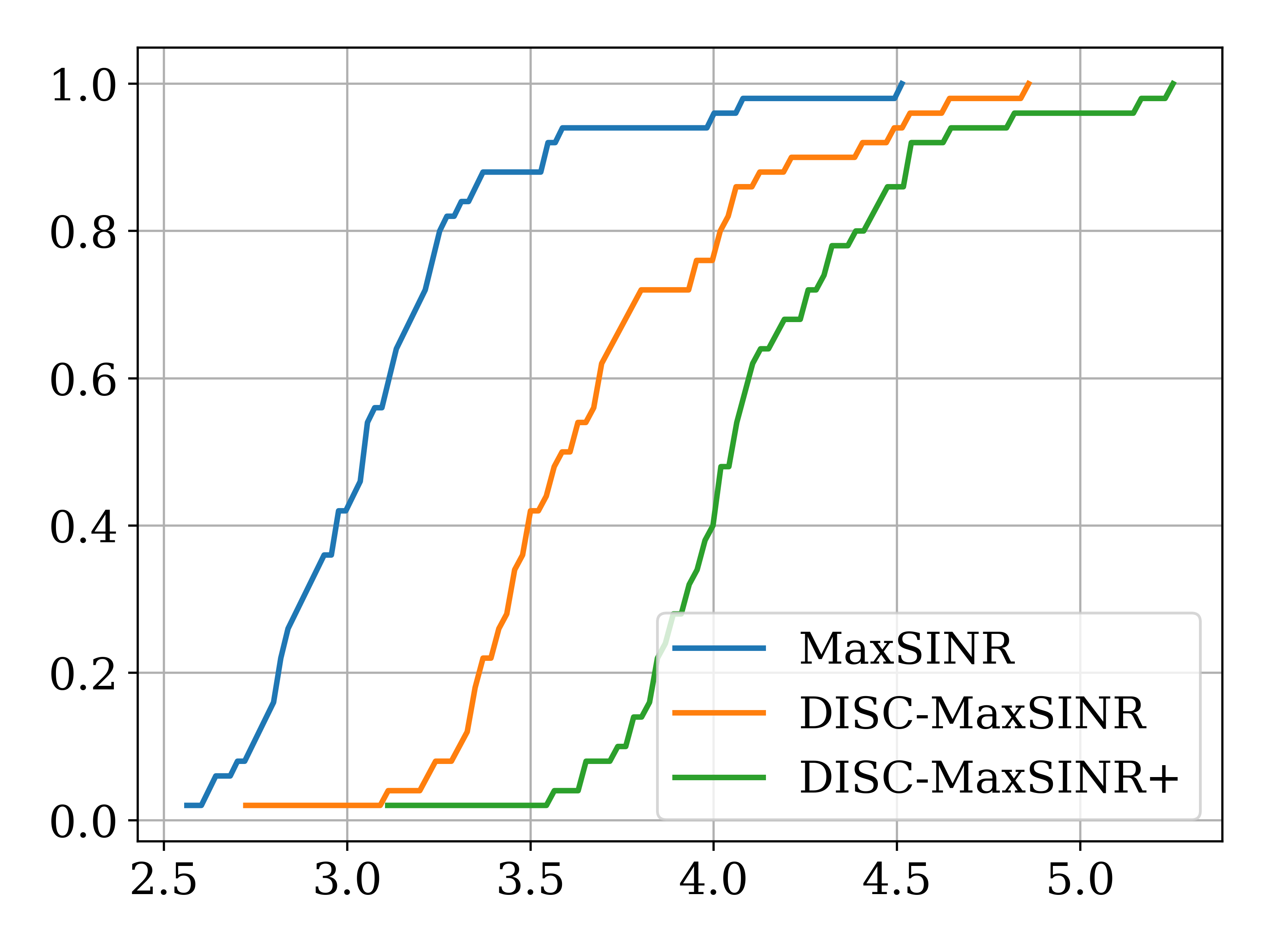}};
        \node[below=of img, node distance=0cm, yshift=1.4cm] {Sumrate (bits/s/Hz)};
        \node[left=of img, node distance=0cm, rotate=90, anchor=center,yshift=-1.1cm]{CDF};
        \node[above=of img, node distance=10cm, yshift=-1.4cm] {$\text{SNR}=12 \text{dB}$, $4-$PAM, $5$ users};
    \end{tikzpicture}
    \caption{Comparison of sumrate between \maxsnospace, \discnospace, and \discpl when $m=4$ and SNR$=12 \text{dB}$ for asymmetric channel settings. The comparison shows that retraining the precoding vectors obtained from \maxs greatly improves the sumrate performance.}
    \label{fig: NN comparison with maxsinr with asymmetric channels with 5 users}
\end{figure}

We consider two specific channel settings to evaluate our approach's performance over the conventional method. Firstly, we consider a symmetric channel setting in~\figgref{fig: NN comparison with maxsinr with symmetric channels} where all the non-diagonal channel entries in the channel matrix have the same value. Secondly, we consider the asymmetric channel settings in~\figgref{fig: NN comparison with maxsinr with asymmetric channels} and~\figgref{fig: NN comparison with maxsinr with asymmetric channels with 5 users} where the channel entries are randomly generated with a fixed interference strength parameter $\bld{\alpha}_{ij}$.

\noindent \textbf{Test setup. \ }
The comparisons were carried out with three different simulation setups. The baseline is the conventional \maxs algorithm, termed as ``\maxsnospace", where input messages are discrete, the encoder and the receiver are derived from the algorithm in~\ref{alg: maxsinr algorithm}. The second baseline is the ``\discnospace" algorithm, which has discrete messages, reuses the \maxs encoder but uses the \ml decoder instead of the linear receiver. The proposed algorithm is the ``\discplnospace" algorithm, which has discrete messages, uses the neural network-based approach to learn the encoder and the power allocation parameters, and uses the \ml decoder. For the test setup, we consider the sumrate is computed from the mutual information between the transmitted and the estimated messages as given in~\eqref{Eqn: sum rate equation} and~\eqref{Eqn: mutual information from probabilities} through the \ml decoding rule in~\eqref{Eqn: Decoding rule for interference network}. The SNRs are chosen to ensure that the interference is neither strong nor weak.

In~\figgref{fig: NN comparison with maxsinr with symmetric channels}, we consider the symmetric channels, for which the channel phase $\bld{\theta}_{ij}$ between the $i^\text{th}$ receiver and the $j^\text{th}$ transmitter is set to a constant $\bld{\theta}_{ij} = \theta$, given by the values on the $x$-axis of the plot. The angles denote the alignment between the desired channel direction and the channel between the interfering users, i.e., $\pi/12$ is in the same direction as the desired signal. In contrast, an angle of $\pi/2$ is orthogonal to the desired signal. The performance of the \disc algorithm is better than the \maxs algorithm as the \ml decoder is able to have non-linear decision boundaries which is better than the linear decoder. The performance of the \discpl algorithm is better than both the baselines, especially when the angle between the desired and the interfering channels (x-axis values) is smaller, denoting that the performance of the neural network approach is better when the interfering signals are stronger. This is because the neural network approach can find better transmit messages not constrained by uniformly placed constellation points of the \maxs algorithm. This enables the transmitted signals from different users to align the interfering signals in an orthogonal direction better than the \maxs algorithm.

In~\figgref{fig: NN comparison with maxsinr with asymmetric channels}, we consider asymmetric channels and show the average performance over $50$ asymmetric channel settings where the channel phase values were chosen randomly between $-\pi/2$ and $\pi/2$. We plot the CDF of the obtained sum rate across the three algorithms. The value on the $y-$ axis gives the probability that the sum rate will be statistically below the corresponding value on the $x-$axis if infinite simulations are run. It shows that more significant improvement is observed for asymmetric channels as the scope of finding a non-linear mapping to avoid multiuser interference is greater.

Furthermore, the performance improvement is more significant when the number of users is increased, as shown in~\figgref{fig: NN comparison with maxsinr with asymmetric channels with 5 users}, where we increased the number of users from three to five. Interestingly, the gap between the \maxs algorithm and the \disc algorithm for the symmetric is more significant than the asymmetric channels. In contrast, for the asymmetric channels, the gap is more significant for the \discpl algorithm than the \disc algorithm. This shows that retraining the encoder improves the sumrate, especially when the channels are asymmetric, and finding a suitable linear beam former for the desired signal is difficult.

\section{Interpretation}
\label{sec: Interpretation}

In the previous sections, we presented the \discpl approach to improve the encoder obtained from the \maxs algorithm via learning. 
We showed that the learned \discpl algorithm outperforms the baselines \disc algorithm, which is the \maxs algorithm modified for discrete messages, and the \maxs algorithm itself. 
Our evaluations demonstrate a significant improvement over the baseline approach, especially for randomly generated asymmetric channel settings.

In this section, we closely examine the learned constellation to shed light on where the improvement comes from. Furthermore, we conduct an ablation study to understand the importance of pretraining in the performance of the \discpl algorithm.

\subsection{Constellation study: \discpl vs. \disc} 
\begin{figure*}
    \centering
    \begin{tikzpicture}[scale = .85, transform shape]
        \node(img){ \includegraphics[width=0.9\textwidth]{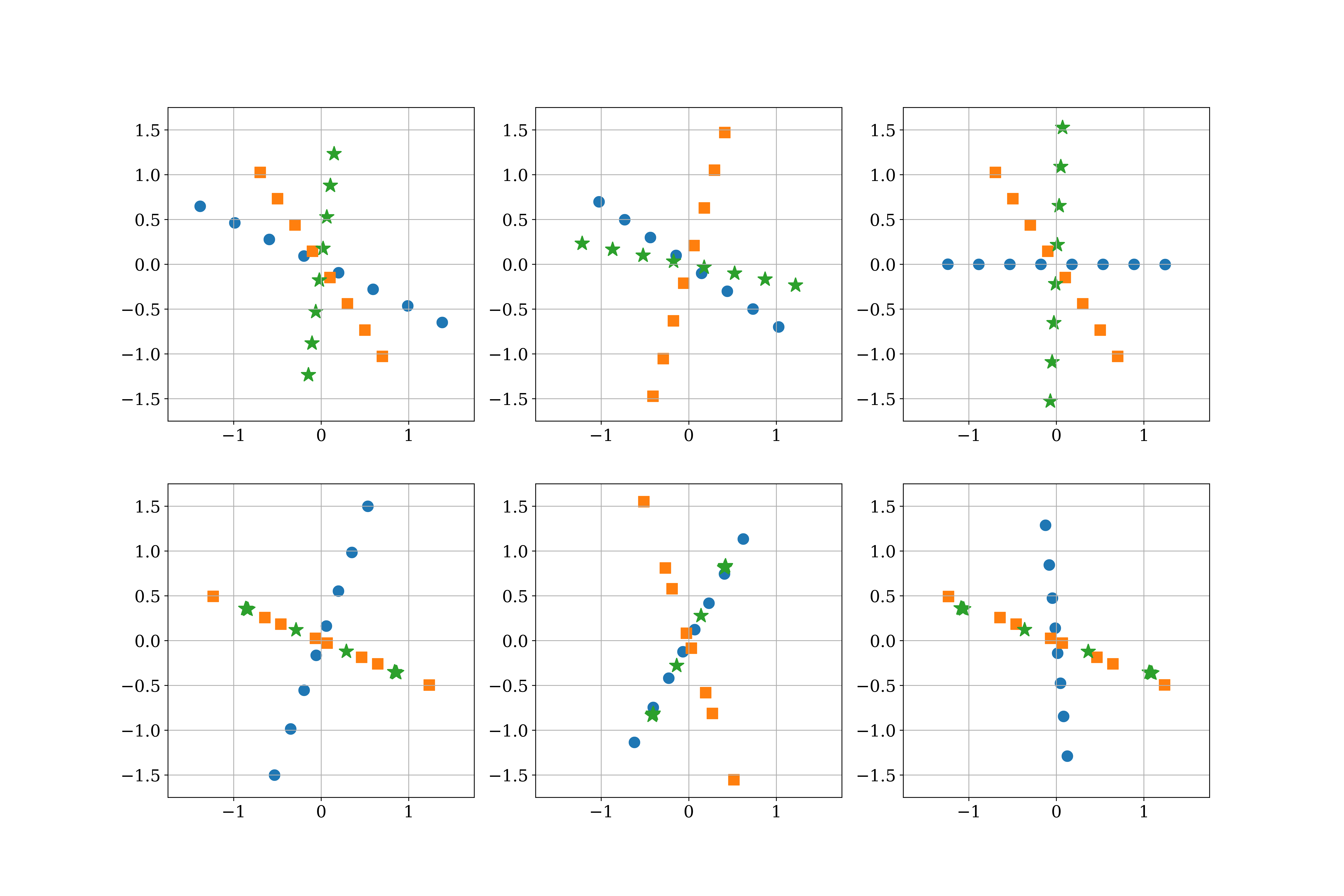}};
        \node[below=of img, node distance=0cm, xshift= -4.4 cm, yshift=2cm, font = \scriptsize] {$\bld{Y}_1(1)$};
        \node[below=of img, node distance=0cm, xshift= 0.2 cm, yshift=2cm, font = \scriptsize] {$\bld{Y}_2(1)$};
        \node[below=of img, node distance=0cm, xshift= 4.8 cm, yshift=2cm, font = \scriptsize] {$\bld{Y}_3(1)$};
        \node[left=of img, node distance=0cm, rotate=90, anchor=center,xshift= 2 cm, yshift=-2.3cm, font = \scriptsize]{$\bld{Y}^{\text{\disc}}_i(2)$};
        \node[left=of img, node distance=0cm, rotate=90, anchor=center,xshift= -2 cm, yshift=-2.3cm, font = \scriptsize]{$\bld{Y}^{\text{\discpl}}_i(2)$};
    \end{tikzpicture}
    \caption{\textit{Received samples:} The plot shows the received samples at the receiver for the \disc (top) and \discpl (bottom) algorithms for the $3$-user interference channel with $8-$PAM messages at $\text{SNR}=18$ dB. The received samples are color-coded (blue circles for messages from user $1$, orange squares for messages from user $2$, and green stars for messages from user $3$).}
    \label{fig: constellation comparison}
\end{figure*}
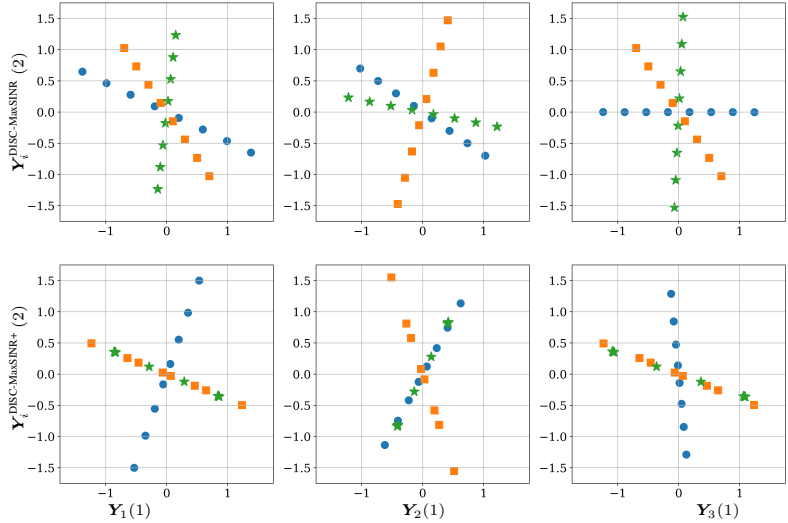

We investigate the constellation learned in the DISC-MaxSINR algorithm and the \discpl algorithm by comparing the orientations of the received samples at each user decoder in relation to the received interfering signals from the other users. This determines if the constellation learned through our approach better aligns the interfering signals orthogonal to our desired signals.

Specifically, in~\figgref{fig: constellation comparison}, we consider an asymmetric channel setting with $3-$users. The interference strength $\bld{\alpha}_{ij}=0.9$ and the channel rotations $\bld{\theta}_{ij}$ selected from the $50$ random realizations considered in~\figgref{fig: NN comparison with maxsinr with asymmetric channels}. The exact values used are provided in the Appendix~\ref{sec: appendix_theta}.    
We plot the received samples from each user before they are summed up at the receivers. Mathematically we plot $\bld{H}_{ij}\bld{X}_j \in \mathbb{R}^2$ for $j=1,2,3$ for the $i$-th receiver at the $i$-th column of~\figgref{fig: constellation comparison} for $i=1,2,3$. 
These received samples are color-coded to show the messages from each user; the circles (blue), squares (orange), and stars (green) correspond to the messages from users $1,2,3$, respectively. The columns correspond to the individual receivers, while each row of the figures corresponds to the algorithms (top: DISC-MaxSINR, bottom: DISC-MaxSINR+). 

\begin{table*}
    \centering
    \begin{tabular}{|c|c|c|c|c|}
    \hline
    \multirow{2}{*}{\textbf{Algorithm}} & \multicolumn{3}{c|}{\textbf{User rate (bits/s/Hz)}} & \multirow{2}{*}{\textbf{Sumrate (bits/s/Hz)}}\\ \cline{2-4}
     & \textbf{User 1} & \textbf{User 2} & \textbf{User 3} & \\ \hline
    \disc & 1.04 & 1.71 & 0.91 & 3.66 \\ \hline
    \discpl & 2.59 & 2.16 & 0.71 & 5.45 \\ \hline
    Improvement  &  1.55& 0.45 & -0.2 & 1.8 \\ \hline
    \end{tabular}
    \caption{Study of the median channel: The table shows the user rates and the sumrate for the \disc and \discpl algorithms for the median channel setting (in terms of improvement).  The improvement in the sumrate is due to the non-linear mapping of the messages to the constellation space. It can be seen that the neural network can sacrifice the third user's rate and obtain a low-rate non-uniform constellation for the second user to improve the overall sumrate.}
    \label{tab: constellation study}
\end{table*}
 
In~\figgref{fig: constellation comparison}, we observe differences between the two constellations (top vs. bottom) in three major aspects: 
(1) the {\em alignment} of the constellation points with respect to other users,  (2) the {\em size} of the constellation given by the number of distinct $PAM$ symbols, and (3) the {\em shape} of the constellation given by the gap between each of these points. We elaborate further on these aspects below.  

1. {\em Alignment.} Ideally, we would like to see the received samples from the interfering users aligned in a perpendicular direction to the desired user. In~\figgref{fig: constellation comparison}, it is evident from the constellation of the first user that \discpl is able to better align the desired symbols with respect to the interfering symbols than DISC-MaxSINR. Consequently, the achievable rate of the first user notably improves, a claim supported by the individual user rates presented in TABLE~\ref*{tab: constellation study}. Specifically, we observe enhancements in the rates of the first and second users with DISC-MaxSINR+, while the rate of the third user experiences a slight reduction. Nonetheless, the overall sum rate is improved.

2. {\em Constellation size.} 
We observe that the \discpl algorithm adjusts the size of the constellation for each user, employing a varying number of distinct constellation points. For instance, while \discpl utilizes 8 distinct blue circle points, it employs only 4 distinct green star points. We conjecture that this modification not only improves the rate of the concerned user but also provides an additional dimension to optimize and results in better interference alignment. 

3. {\em Constellation shape. }
We observe changes in the shape of the constellation, as indicated by the spacing between constellation points for different users. Specifically, the second user's constellation points exhibit non-uniform distances between them.

In the following section, we will study the impact of each of these factors on the sumrate.

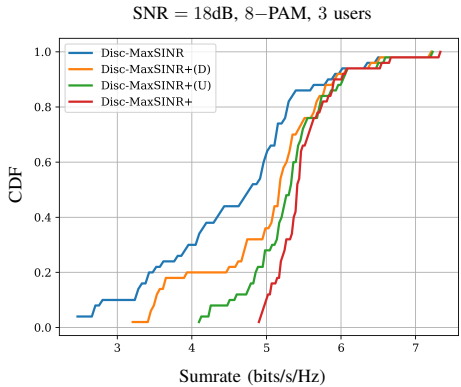
\begin{figure}
    \centering
    \begin{tikzpicture}[scale = .7, transform shape]
        \node(img){\includegraphics[width=0.6\textwidth]{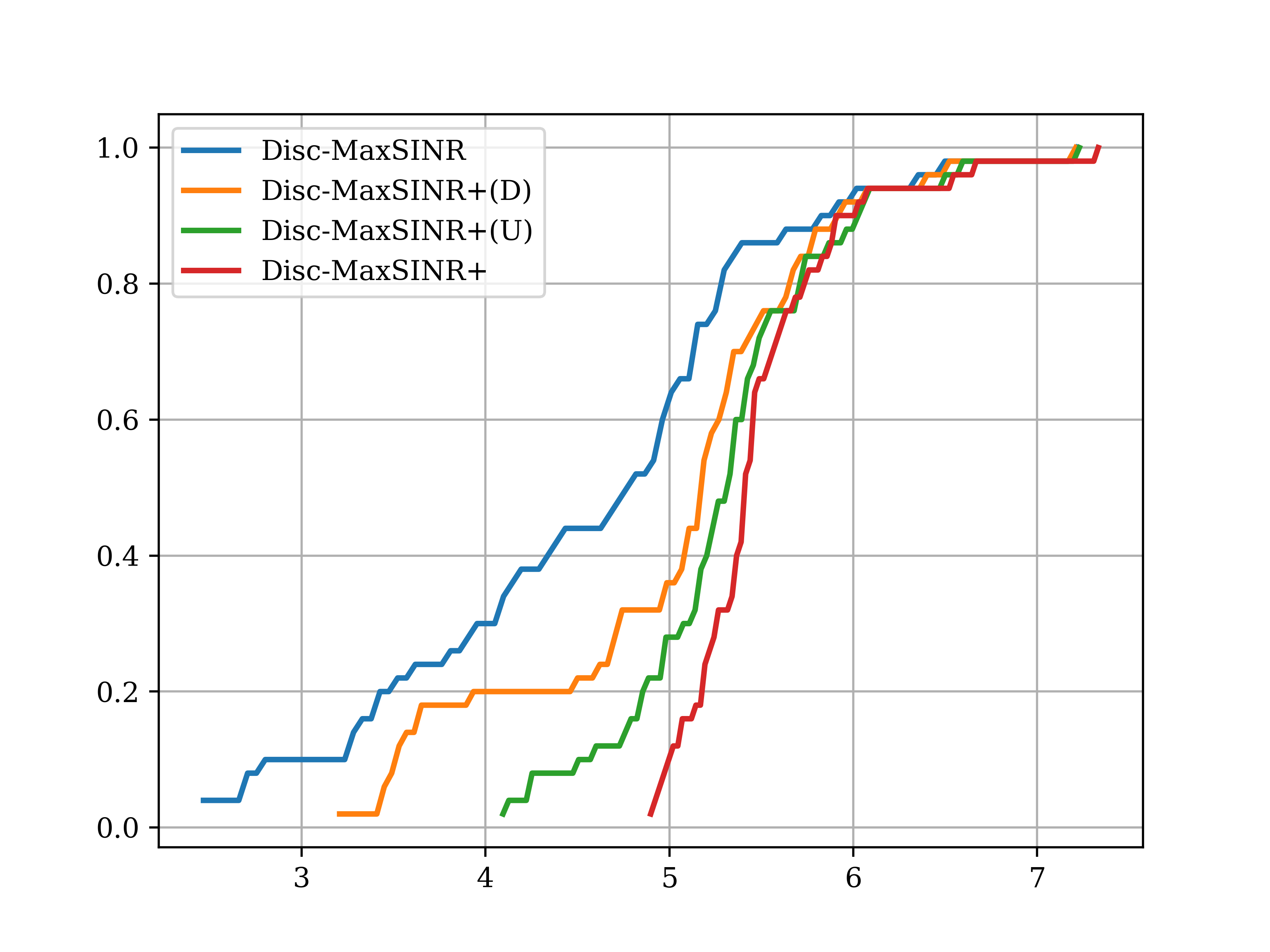}};
        \node[below=of img, node distance=0cm, yshift=1.4cm] {Sumrate (bits/s/Hz)};
        \node[left=of img, node distance=0cm, rotate=90, anchor=center,yshift=-1.5cm]{CDF};
        \node[above=of img, node distance=10cm, yshift=-1.8cm] {$\text{SNR}=18 \text{dB}$, $8-$PAM, $3$ users};
    \end{tikzpicture}

    \caption{The plot shows the comparison of the sumrate CDF of different versions of \discpl algorithm with the \disc algorithm for the $3$-user interference channel with $8-$PAM messages at $\text{SNR}=18$ dB. \discpl(D) represents the \discpl algorithm with only the alignment optimized. \discpl(U) represents the algorithm with the optimized precoding vectors and the size of the constellation as trainable variables. The gaps between these four plots demonstrate the importance of the learned constellation size, shape, and alignment on the enhanced sumrate.}
    \label{fig: rate adaptation ablation study}
\end{figure}

\subsection{Impact of learned constellation size, shape and alignment on the enhanced sumrate}

In this section, we run several ablation studies to investigate the effect of the constellation's size, shape, and alignment on the enhanced sumrate. We implement and compare four algorithms: 

\begin{itemize}[leftmargin=*]
    \item \discpl refers to the learned interference algorithm with learned alignment, adjusted constellation size, and optimized non-uniform constellation. 
    \item \discplnospace(U): We take the fully learned constellation from \discpl algorithm but rearrange the symbol locations such that they are {\em uniformly} spaced. 
    \item \discplnospace(D): We take the fully learned constellation from \discpl algorithm but rearrange the symbol locations such that they are {\em uniformly} spaced and the constellation size is {\em fixed} and constant ($8-$PAM) across the users.
    \item DISC-MaxSINR refers to the non-learning baseline algorithm. 
\end{itemize}

In~\figgref{fig: rate adaptation ablation study}, we run these four algorithms for the set of $50$ channels as in~\figgref{fig: NN comparison with maxsinr with asymmetric channels} and plot their sumrate CDF. From these results, we conclude that the enhanced sum rate is influenced by all three key factors: the learned alignment, adaptive constellation size, and non-uniform constellation shape. 
We also infer that the size of the constellation plays a major role in finding a suitable alignment that helps in interference mitigation. The shape of the constellation also improves the overall sumrate and results in superior performance, and having both these aspects as trainable variables is crucial for the performance of the \discpl algorithm.

\subsection{Ablation study: Importance of Pretraining}
\label{sec: ablation study for pretrained constellation}
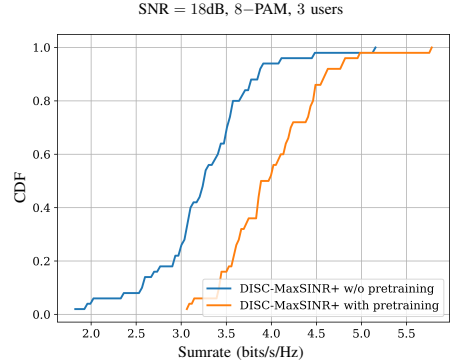
\begin{figure}
    \centering
    \begin{tikzpicture}[scale = .6, transform shape]
        \node(img){\includegraphics[width=0.6\textwidth]{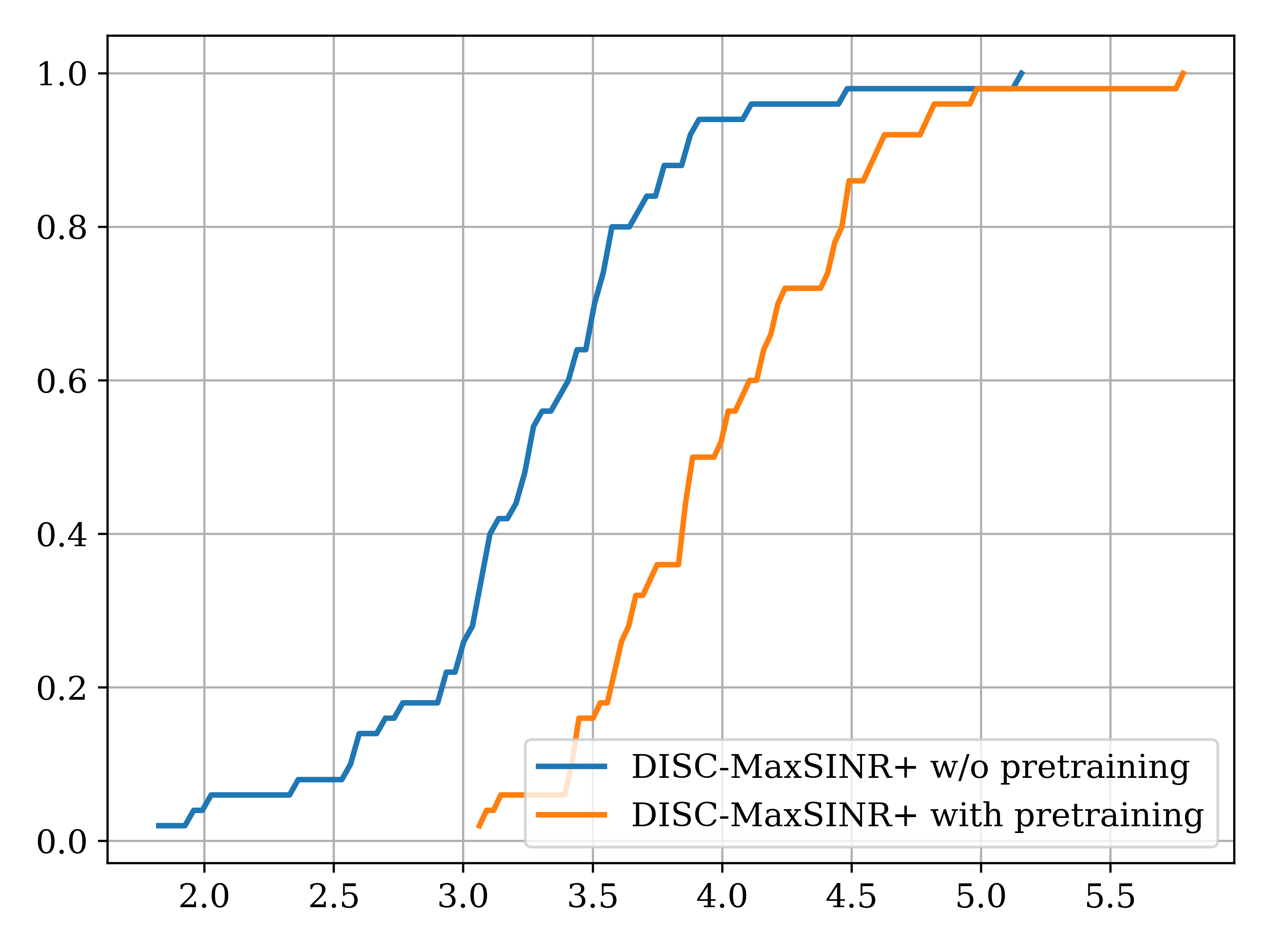}};
        \node[below=of img, node distance=0cm, yshift=1.4cm] {Sumrate (bits/s/Hz)};
        \node[left=of img, node distance=0cm, rotate=90, anchor=center,yshift=-1.1cm]{CDF};
        \node[above=of img, node distance=10cm, yshift=-1.1cm] {$\text{SNR}=18 \text{dB}$, $8-$PAM, $3$ users};
    \end{tikzpicture}

    \caption{Power of pretraining: The plot shows the difference in the sumrate if the network is learned for the same amount of time but with and without pretraining. It clearly shows that the pretraining helps in faster convergence and results in higher sumrates.}
    \label{fig: pretraining comparison}
\end{figure}

In this section, we empirically demonstrate that pretraining actually helps in learning a better encoder. In~\figgref{fig: pretraining comparison}, we consider a series $50$ random channels which were trained for the same setting for the same number of epochs but with different initializations. In one case, we consider random initializations, where the precoding vectors were randomly assigned, while in the other case, the precoding vectors were initialized to the vectors obtained from \maxs algorithm. It is evident that the pretraining helps in improving the sum rate. The objective function representing the optimization of the encoder to maximize the sum rate is non-convex and the pretraining helps in avoiding local minima. 

\section{Conclusion}
\label{sec: Conclusion}

Theoretical studies of interference alignment have focused largely on high SNR asymptotics (such as degrees of freedom) which lead to beamforming solutions with Gaussian signaling, as embodied in the MaxSINR algorithm. In practical regimes of interest, however, SNR is only moderately high, power control is important, and the size and shape of discrete constellations at each user play a crucial role in the relative alignments of desired and interfering signals. Given the difficulty of analytically finding the optimally aligned discrete constellations, we investigate the efficacy of deep learning approaches for this task. In particular, the combined utility of MaxSINR solutions and deep neural networks is our focus. Incorporating insights from MaxSINR solutions as domain knowledge to initialize the neural networks, we find faster convergence to novel constellations that offer improved alignments not only in terms of beamforming directions but also in terms of the effective constellation at the receiver, resulting in better sum-rate performance. The optimistic picture from this work paves the way for further studies to distill new design principles for discrete interference alignment.

\bibliographystyle{IEEEtran}
\bibliography{References}

\appendix
\section{Appendix I}
\label{sec: appendix_theta}
The values for channel rotation $\bld{\theta}_{ij}$ used in plotting the constellation in~\figgref{fig: constellation comparison}(Section~\ref{sec: Interpretation}) is given as\[
\begin{bmatrix}
$0$ & $-49.81$ & $4.17$ \\
$9.13$ & 0 & $-81.79$\\
$-25.07$ & $-49.84$ & $0$
\end{bmatrix}
\]

\end{document}